\documentclass[12pt]{article}
\usepackage{a4wide}
\usepackage{amsmath,amssymb,bbm}
\usepackage{multirow}
\usepackage{comment}
\usepackage{pstricks}
\usepackage{graphicx}

\newsavebox{\uuunit}
\sbox{\uuunit}
    {\setlength{\unitlength}{0.825em}
     \begin{picture}(0.6,0.7)
        \thinlines
        \put(0,0){\line(1,0){0.5}}
        \put(0.15,0){\line(0,1){0.7}}
        \put(0.35,0){\line(0,1){0.8}}
       \multiput(0.3,0.8)(-0.04,-0.02){12}{\rule{0.5pt}{0.5pt}}
     \end {picture}}
\newcommand {\unity}{\mathord{\!\usebox{\uuunit}}}

\begin{document}

\begin{titlepage}
\begin{center}

\hfill UG-13-12

\vskip 1.5cm

{\LARGE \bf  Branes, Weights and Central Charges}

\vskip 1cm

{\bf Eric A.~Bergshoeff\,$^1$, Fabio Riccioni\,$^2$  and Luca Romano\,$^3$}

\vskip 25pt

{\em $^1$ \hskip -.1truecm Centre for Theoretical Physics,
University of Groningen, \\ Nijenborgh 4, 9747 AG Groningen, The
Netherlands \vskip 5pt }

{email: {\tt E.A.Bergshoeff@rug.nl}} \\

\vskip 15pt

{\em $^2$ \hskip -.1truecm
 INFN Sezione di Roma,   Dipartimento di Fisica, Universit\`a di Roma ``La Sapienza'',\\ Piazzale Aldo Moro 2, 00185 Roma, Italy
 \vskip 5pt }

{email: {\tt Fabio.Riccioni@roma1.infn.it}} \\

\vskip 15pt

{\em $^3$ \hskip -.1truecm  Dipartimento di Fisica and INFN Sezione di Roma, Universit\`a di Roma ``La Sapienza'',\\ Piazzale Aldo Moro 2, 00185 Roma, Italy
 \vskip 5pt }

{email: {\tt  Luca.Romano@roma1.infn.it}} \\

\end{center}

\vskip 0.5cm

\begin{center} {\bf ABSTRACT}\\[3ex]
\end{center}

We study the properties of half-supersymmetric branes in string theory with 32 supercharges from a purely group-theoretical point of view using the U-duality symmetry of maximal supergravity and the R-symmetry of the corresponding supersymmetry algebra. In particular, we show that half-supersymmetric branes are always associated to the longest weights of the U-duality representation of the potentials coupling to them. We compare the features of branes with three or more transverse directions (that we call ``standard'' branes) to those with two or less transverse directions (that we denominate ``non-standard'' branes).  We show why the BPS condition of the  non-standard branes  is in general degenerate and for each case we calculate this degeneracy. We furthermore show how the orbits of multi-charge configurations of non-standard branes can be calculated and give the U-duality invariants describing these orbits.
We show that different orbits of non-standard branes can have the same BPS condition.

\end{titlepage}

\newpage
\setcounter{page}{1} \tableofcontents

\newpage

\setcounter{page}{1} \numberwithin{equation}{section}

\section{Introduction}

It is since long known that ``branes'', i.e.~massive objects with a number of worldvolume and transverse directions, play a crucial role in string theory and M-theory. Historically, the first example of a brane other than a string was the
eleven-dimensional supermembrane \cite{Bergshoeff:1987cm}. An important class of branes are the Dirichlet branes or, shortly,
D-branes of ten-dimensional superstring theory \cite{Polchinski:1995mt}. These branes are non-perturbative in the sense that
their brane tension scales with the inverse of the string coupling constant. D-branes played a decisive role in the
calculation of the entropy of a certain class of black holes \cite{Strominger:1996sh}. Branes also play a central role in the AdS/CFT correspondence \cite{Maldacena:1997re} and the brane-world scenario \cite{Randall:1999ee}.

Much information about branes in string theory and/or M-theory can be obtained by studying the low-energy
approximation of these theories which is a supergravity theory that realizes the gauging of a specific supersymmetry algebra.
For instance, the mere fact that eleven-dimensional supergravity contains a 3-form potential is already indicative of the
fact that M-theory contains a membrane since 3-forms naturally couple to membranes. The fact that this membrane is actually
a supermembrane which breaks half of the supersymmetry follows from the construction of a kappa-symmetric
supermembrane action \cite{Bergshoeff:1987cm}. The occurrence of an eleven-dimensional supermembrane can also be deduced
from the presence of a 2-form central charge in the eleven-dimensional super-Poincar\'e algebra \cite{de Azcarraga:1989gm}.

Due to its relevance it is important to classify the branes of string theory and/or M-theory. One way to do this is to scan
the possible $(p+1)$-forms in supergravity and verify whether they may couple to a supersymmetric brane by investigating the corresponding kappa-symmetric worldvolume action. In the case of $D$-dimensional supergravity with maximal supersymmetry such an investigation has been done for all $(p+1)$-forms with $0 \le p \le D-4$. One finds that to each  $(p+1)$-form potential there corresponds precisely one half-supersymmetric $p$-brane. In the case that the potential transforms according to a certain representation of the U-duality group one finds as many half-supersymmetric branes as the dimension of that U-duality representation.

One may wonder, given the above result, what more information about branes can be extracted from the low-energy supergravity theory. The reason why more information can be extracted is that our knowledge about the general structure of a supergravity theory has considerably been improved in recent years. Up to some years ago most of our knowledge about the $(p+1)$-forms of supergravity was restricted to the ones with $0\le p \le D-4$. Note that all such $p$-forms describe physical
degrees freedom of the supergravity multiplet and that  some potentials are related to each other by electromagnetic
duality\,\footnote{
In general a $(p+1)$-form potential in $D$ dimensions is dual to a $(D-p-3)$-form potential.}. A common feature of these potentials is that they all couple to a brane whose number of transverse directions is more than or equal to three. Such branes approach flat Minkowski spacetime asymptotically
and have a finite energy density. We will call such branes ```standard'' branes.

In this work we will focus on  the branes that have less than three transverse directions and
compare them with the standard branes.
These so-called ``non-standard'' branes  couple to $(p+1)$-form potentials with $p=D-3, p=D-2$ or $p=D-1$. A special class is formed by the $(D-2)$-form potentials of supergravity. These potentials are special in the sense that they
are dual to 0-form potentials, or scalars, but the duality relations do not imply that the number of $(D-2)$-form potentials is equal to the number of scalars. The $(D-2)$-form potentials couple to so-called ``defect branes'', i.e.~branes with two transverse directions. In four dimensions they occur as cosmic strings \cite{Greene:1989ya} while in
ten dimensions they are the seven-branes \cite{Gibbons:1995vg} that underly F-theory \cite{Vafa:1996xn}.
Defect branes differ from standard branes in the sense that they are not asymptotically flat and cannot be given finite energy unless one takes several of them in the presence of an orientifold. Another notewearthy feature is that the number of $(D-2)$-form potentials is not equal to the number of half-supersymmetric $(D-3)$-branes \cite{Bergshoeff:2011se}.
This result is based on an analysis of the Wess-Zumino (WZ) terms in the world-volume action of a single $(D-3)$-brane,
see, e.g., \cite{Bergshoeff:2010xc}. Based on U-duality arguments we know that  for those cases that a gauge-invariant  WZ
term, consistent with world-volume supersymmetry,  can be constructed a kappa-symmetric worldvolume action exists. Furthermore, we expect that configurations of $(D-3)$-branes with a finite energy, using the same techniques as in ten dimensions, can be constructed.

It is natural to extend the discussion of the non-standard $(D-3)$-branes to the non-standard branes with one and zero
transverse directions. Such branes are called ``domain walls'' and ``space-filling branes'', respectively. Domain walls
play an important role in the AdS/CFT correspondence since they describe the renormalization group flow of the boundary conformal field theory. Space-filling branes are used in string theory to define strings with sixteen supercharges.
Domain walls and space-filling branes are
even more special than the defect branes in the sense that they couple to potentials that do not describe any physical degree of freedom in the corresponding supergravity theory\,\footnote{Note that the $(D-1)$-form potentials that couple to domain walls are dual
to an integration constant such as a gauge coupling constant or a mass parameter.}. Much less was known about these $(D-1)$-form and $D$-form potentials because,
 unlike the $(p+1)$-form potentials with $0\le p\le D-4$, their existence does not follow from the representation theory of the
 supersymmetry algebra.

One of the remarkable developments about our knowledge on supergravity in recent years has been that a full classification
has been given  of all $(D-1)$-form and $D$-form potentials that can
be added to maximal supergravity. This has been achieved using three different techniques. By an explicit verification of the supersymmetry algebra it was shown that IIA and IIB supergravity allow such potentials and a classification, including the U-duality representations in the case of IIB supergravity, was given \cite{Bergshoeff:2005ac}. Although in principle possible, it is very impractical to extend this method to all lower dimensions. Fortunately, it turns out that a full classification for all dimensions $3\le D\le 11$ can be given
\cite{Riccioni:2007au,Bergshoeff:2007qi} making use of the properties of the
very extended Kac-Moody algebra $E_{11}$  \cite{West:2001as}.
Remarkably, independently a full classification, including all dimensions lower than ten, has been
given using the so-called embedding tensor technique \cite{deWit:2008ta}.

Given the $(p+1)$-forms and their U-duality representations the next question to answer is how many components of these
U-duality representations correspond to half-supersymmetric $p$-branes. For the standard branes the answer is simple: each component of the U-duality representation corresponds to a half-supersymmetric brane.  However, for the half-supersymmetric
non-standard  branes
the answer is less clear. Demanding that a gauge-invariant WZ term can be constructed, consistent with worldvolume
supersymmetry, the half-supersymmetric non-standard branes of maximal supergravity have been classified in our earlier work
\cite{Bergshoeff:2010xc,Bergshoeff:2011qk,Bergshoeff:2012ex}. An alternative derivation, based upon the counting of the real roots of the very extended Kac-Moody algebra $E_{11}$, has been given in \cite{Kleinschmidt:2011vu}.

It is the purpose of this work to give a simple and elegant group-theoretical explanation of why the ``WZ method'' of  \cite{Bergshoeff:2010xc,Bergshoeff:2011qk,Bergshoeff:2012ex} and the ``real-root method'' of \cite{Kleinschmidt:2011vu} give the same  result. In general, given a supergravity theory with a $(p+1)$-form potential  in a specific U-duality representation, the half-supersymmetric branes resulting from the WZ-term analysis correspond to the weights that can be chosen as highest weights of that U-duality representation\,\footnote{If, for given $p$, there
are several irreducible U-duality representations, the half-supersymmetric branes belong to the highest-dimensional representation.}. A U-duality representation has typically weights of different lengths, and the weights that can be chosen as highest weights are those of maximum length. This simple observation leads to a way of counting the half-supersymmetric branes by counting the longest weights of the corresponding U-duality representation.
As will be better explained in the conclusions, the longest weights of the U-duality representation of a field corresponding to a brane precisely correspond to the real roots of $E_{11}$.

The above ``longest-weight rule'' explains several properties of the standard and non-standard branes we already mentioned.
It turns out that all $(p+1)$-forms that couple to the standard branes only occur in U-duality representations where
all weights  have equal length and hence are longest weights. This explains why for standard branes each component of the U-duality representation corresponds to a half-supersymmetric  brane. The U-duality representations corresponding to
$(p+1)$-forms that couple to the non-standard branes are different: they have weights of different lengths and only the longest weights correspond to the half-supersymmetric non-standard branes. Such representations have the defining property that they contain more than one  so-called {\sl dominant weight}, a notion that we will explain in the main text of this work.
An interesting special case is formed by the
$(p+1)$-forms that couple to the defect branes. These $(p+1)$-forms are always in the adjoint representation of the U-duality group $G$. These representations have the property that all weights are longest weights except for the Cartan generators.
This explains the result of \cite{Bergshoeff:2011se} that out of the $\text{dim}\, G$ $(p+1)$-forms that couple to
the defect branes only $\text{dim}\, G - \text{rank}\, G$ components couple to  half-supersymmetric defect branes. For instance, IIB supergravity has three 8-form potentials that transform as the  ${\bf 3}$ of SL(2,$\mathbb{R}$). Only two of them
couple to a half-supersymmetric defect brane: the D7-brane and its S-dual.

There is a second  crucial difference between standard and non-standard branes: while for standard branes there is a one-to-one relation between half-supersymmetric branes and the BPS conditions they satisfy, in the case of non-standard branes this relation is many-to-one, i.e.~several non-standard branes may satisfy the same  BPS condition~\cite{Bergshoeff:2011se,Bergshoeff:2012pm,Bergshoeff:2012jb}. This implies that, unlike the
standard branes, the non-standard  branes may form bound states that satisfy the same BPS condition. In this work we
will explain why this property holds from a purely group-theoretical point of view by comparing the U-duality representations
of the $(p+1)$-forms with the $R$-symmetry representations of the central charges in the supersymmetry algebra. In this way
we are able to derive the explicit degeneracies of the different BPS conditions, i.e.~how many branes satisfy the
same BPS condition.

In this work we will point out a third difference between the behaviour of the standard and non-standard branes which concerns the brane orbits. Given a half-supersymmetric brane one can consider its orbit under the action of the U-duality symmetry group.
 All half-supersymmetric branes in maximal supergravity define highest-weight orbits.  These highest-weight orbits are single-charge orbits. In the case of standard branes it has been shown that, if not all longest weights can be reached from the initial configuration by an infinitesimal transformation of the group $G$ (that is a transformation generated by the corresponding Lie algebra $g$), one can
consider a two-charge state that is the sum of the initial state and  one that cannot be reached by the initial state. One can then compute the orbit of this 2-charge configuration. In case the  single-charge
and two-charge configurations do not fill up the full U-duality representation one continues to
consider three-charge configurations etc. This procedure can be iterated until one has a configuration in which all the weights can be reached \cite{Lu:1997bg}. In \cite{Bergshoeff:2012ex} we applied this method to compute the single-charge orbits for all the non-standard branes. In this work we will show how the multi-charge orbits of the non-standard branes can be calculated as well. A crucial difference with the standard brane orbits will be the existence of half-supersymmetric multi-charge orbits.  We will furthermore show how the different standard and non-standard brane orbits can be characterized in terms of invariants of the U-duality group \cite{Ferrara:1997ci}.

This work is organized as follows. In section 2 we show the relation between half-supersymmetric branes and the longest weights of the U-duality representation of the $(p+1)$-forms that couple to these branes. In particular, we will clarify the longest-weight rule mentioned earlier and use it to explain the number of standard and non-standard $(p+1)$-branes as compared to the number of U-duality components of the  $(p+1)$-form potentials.  Next, in section 3 we focus on a second difference between standard and non-standard branes which concerns the supersymmetry properties. More prescisely, we discuss the
relation between the BPS conditions and the central charges in the supersymmetry algebra and calculate the degeneracies of the different BPS conditions. We will show that, unlike the standard banes, different non-standard branes may satisfy the same BPS condition.
Finally, in section 4 we discuss the difference between the standard-brane and non-standard-brane orbits.
We first review the standard-brane orbits and next show how to compute the orbits of the  non-standard branes
including the multi-charge orbits. We furthermore give the U-duality invariant that characterizes the different orbits. Our conclusions are presented in the last section.

\section{Weights of half-supersymmetric branes}

In this section  we will show that
the potentials associated to standard branes belong to irreducible representations with only one dominant weight, which is the highest weight of the representation, while
the potentials associated to non-standard branes belong to irreducible representations with more than one dominant weight. If a representation contains more than one dominant weight, each dominant weight other than the highest weight defines a sub-representation whose weights are shorter than the highest weight, while if a representation has only one dominant weight, this means that all the weights have the same length.
We will show that all half-supersymmetric branes  correspond to the longest weights in the irreducible representation of the potential. In particular, this explains why  the number of standard branes always coincides with the dimension of the corresponding representation, while the number of non-standard branes is less than the dimension of the corresponding representation.
In order to make all these statements clear, we will give in the first part of this section a review of the Lie algebra tools that are needed to understand them\,\footnote{For a pedagogical introduction to Lie algebras, see e.g. \cite{Cahn:1985wk}.}. In the second part of this section we will  proceed with identifying the branes with the longest weights within each irreducible representation in any dimension.

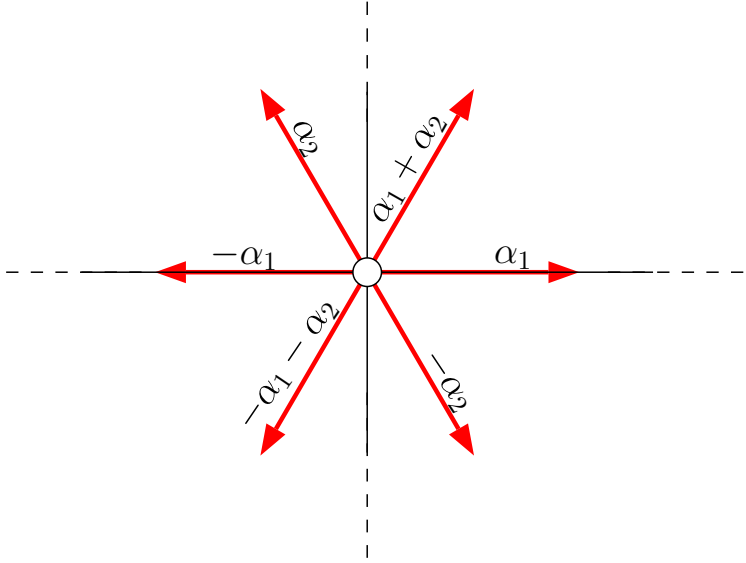
\begin{figure}
 \centering
\scalebox{1} 
{
\begin{pspicture}(0,-3.8992147)(9.81,3.9364839)
\psline[linewidth=0.06cm,arrowsize=0.05291667cm
4.0,arrowlength=1.4,arrowinset=0.0,linecolor=red]{->}(4.8,-0.08921477)(3.38,2.3507853)
\psline[linewidth=0.06cm,arrowsize=0.05291667cm
4.0,arrowlength=1.4,arrowinset=0.0,linecolor=red]{->}(4.8,-0.08921477)(1.98,-0.08921477)
\psline[linewidth=0.06cm,arrowsize=0.05291667cm
4.0,arrowlength=1.4,arrowinset=0.0,linecolor=red]{->}(4.8,-0.08921477)(6.22,2.3507853)
\psline[linewidth=0.06cm,arrowsize=0.05291667cm
4.0,arrowlength=1.4,arrowinset=0.0,linecolor=red]{->}(4.8,-0.08921477)(3.38,-2.5292149)
\psline[linewidth=0.06cm,arrowsize=0.05291667cm
4.0,arrowlength=1.4,arrowinset=0.0,linecolor=red]{->}(4.8,-0.08921477)(7.62,-0.08921477)
\psline[linewidth=0.06cm,arrowsize=0.05291667cm
4.0,arrowlength=1.4,arrowinset=0.0,linecolor=red]{->}(4.8,-0.08921477)(6.22,-2.5292149)
\usefont{T1}{ppl}{m}{n}
\rput{55.888874}(0.58888847,-3.6979384){\rput(3.7434375,-1.2742147){\large $-\alpha_{1}-\alpha_{2}$}}
\usefont{T1}{ppl}{m}{n}
\rput{60.39754}(3.8652787,-3.9893463){\rput(5.3234377,1.3457853){\large $\alpha_{1}+\alpha_{2}$}}
\psline[linewidth=0.02cm](1.0,-0.08921477)(8.6,-0.08921477)
\psline[linewidth=0.02cm](4.8,2.3107853)(4.8,-2.4892147)
\psline[linewidth=0.02cm,linestyle=dashed,dash=0.16cm 0.16cm](4.8,-3.8892148)(4.8,-1.6892148)
\psline[linewidth=0.02cm,linestyle=dashed,dash=0.16cm 0.16cm](4.8,1.9107852)(4.8,3.5107853)
\psline[linewidth=0.02cm,linestyle=dashed,dash=0.16cm 0.16cm](0.0,-0.08921477)(1.8,-0.08921477)
\psline[linewidth=0.02cm,linestyle=dashed,dash=0.16cm 0.16cm](8.0,-0.08921477)(9.8,-0.08921477)
\usefont{T1}{ppl}{m}{n}
\rput(6.7434373,0.10578523){\large $\alpha_{1}$}
\usefont{T1}{ppl}{m}{n}
\rput{-56.242115}(3.9211056,4.189283){\rput(5.8434377,-1.5542147){\large $-\alpha_{2}$}}
\usefont{T1}{ppl}{m}{n}
\rput(3.1634376,0.12578523){\large $-\alpha_{1}$}
\usefont{T1}{ppl}{m}{n}
\rput{-59.27895}(0.5316899,4.2266393){\rput(3.9434376,1.6657852){\large $\alpha_{2}$}}
\pscircle[linewidth=0.02,dimen=outer,fillstyle=solid](4.8,-0.08921477){0.2}
\end{pspicture}
}
\caption{The roots of the Lie algebra $sl(3)$.  The roots are painted in red because they are the six longest weights of the ${\bf 8}$. In general, for any $sl(3)$ representation, we paint in red the longest weights of the representation.  \label{rootssl3}}
\end{figure}

The simple Lie algebra $sl(2)$ is the prototype of any simple finite-dimensional Lie algebra. The generators of $sl(2)$ are the Cartan generator $L_3$ and the creation and annihilation operators $L_+$ and $L_-$. The commutator between the Cartan generator $L_3$ and the $L_\pm$ generators is given by
  \begin{equation}
   [ L_3 , L_\pm ] = \pm  L_\pm \quad .
\end{equation}
Similarly, for any simple Lie algebra $g$ of dimension $d$ and with Cartan subalgebra $h$ of dimension $r$, the $d-r$ generators which are not Cartan can be split into $(d-r)/2$ creation operators $E_\alpha$ and $(d-r)/2$ annihilation operators $E_{-\alpha}$ obeying the commutation relations
  \begin{equation}
  [ H , E_{\pm \alpha} ] = \pm \alpha (H) E_{\pm \alpha}
\end{equation}
with the Cartan generators $H \in h$, where the  roots $\pm \alpha (H)$ are linear functions of $H$\,\footnote{\label{positiverootfootnote}One defines $\alpha (H)$ as the {\it positive} roots, and their opposite as the {\it negative} ones. Clearly, this definition corresponds to the choice of which operators are creation operators and which ones are annihilation operators. We will make this more clear later.}.
Moreover, for every $E_\alpha$ there is a corresponding $H_\alpha$ such that the root $\alpha (H)$ is proportional to the Cartan-Killing form $(H_\alpha , H)$. Thus, the Cartan-Killing form induces a scalar product $< \alpha ,\beta >$ in the space of roots, which is proportional to $(H_\alpha , H_\beta )$.  One can then associate to each root a vector in an 
$r$-dimensional vector space with Euclidean signature. One then defines the {\it simple} roots $\alpha_1 , ..., \alpha_r$ as those such that all the other positive roots (see footnote~$^{\ref{positiverootfootnote}}$) can be obtained as positive sums of them. We consider as a simple example the roots of $sl(3)$, which are drawn in Fig.~\ref{rootssl3}. The simple roots are $\alpha_1$ and $\alpha_2$, while the other positive root is their sum $ \alpha_1+ \alpha_2$.
 Actually, in the diagram any pair of roots that form an angle of $2\pi /3$ can be chosen as simple roots. The choice  made in Fig.~\ref{rootssl3} defines the operators $E_{\alpha_1}$, $E_{\alpha_2}$ and $E_{\alpha_1 +\alpha_2}$ as ``creation'' operators, and correspondingly $E_{-\alpha_1}$, $E_{-\alpha_2}$ and $E_{-\alpha_1 -\alpha_2}$ are ``annihilation'' operators.

In $sl(2)$, irreducible representations are labelled by $j_{\rm max}$ (which takes integer or half-integer positive  values), which is the eigenvalue of $L_3$ with eigenvector ${\bf \Lambda}_{ j_{\rm max}}$ annihilated by $L_+$. Acting with $L_-$, one lowers by 1 the $L_3$ eigenvalue.  Proceeding this way, one can lower the eigenvalue down to $-j_{\rm max}$, whose corresponding eigenvector ${\bf \Lambda}_{ - j_{\rm max}} $ is annihilated by $L_-$. This altogether forms a representation of dimension $2j_{\rm max} +1$. Analogously, in a generic simple Lie algebra $g$, irreducible representations are labelled by eigenstates of the Cartan generators (i.e.~weight vectors) ${\bf \Lambda}_{ W_{\rm max} }$ of eigenvalue (weight) $W_{\rm max}(H)$, such that $E_{\alpha_i} {\bf  \Lambda}_{W_{\rm max} } =0$ for all simple roots $\alpha_i$\,\footnote{This implies that $E_{\alpha} {\bf  \Lambda}_{W_{\rm max} } =0$ vanishes for all positive roots $\alpha$.}. Such weights are called {\it highest weights}. Acting with $E_{-\alpha_i}$ on ${\bf  \Lambda}_{W_{\rm max} }$, one either gets zero or a weight vector ${\bf  \Lambda}_{W_{\rm max} - \alpha_i }$ of eigenvalue  $W_{\rm max}(H) - \alpha_i (H)$. One can then keep acting with
$E_{-\alpha_i}$ until one finds a $q_i$ such that
 \begin{equation}
 ( E_{-\alpha_i} )^{q_i+1} {\bf \Lambda}_{W_{\rm max} } =0 \ .
 \end{equation}

Exactly as for the roots, for every weight vector ${\bf \Lambda}_{W } $ there is a corresponding Cartan generator $H_W$ such that the weight $W (H)$ is proportional to the Cartan-Killing form $(H_W , H)$. Thus,  one can define a scalar product $<W, \alpha >$ between the weight and the roots, and  draw the weight  on the $r$-dimensional vector space of the roots. In terms of the scalar product, $q_i$ is then given by the relation
  \begin{equation}
  q_i = \frac{2 <W_{\rm max} , \alpha_i > }{<\alpha_i , \alpha_i >}\quad ,\label{qihighestweight}
\end{equation}
where clearly the $q_i$'s must be non-negative for consistency.
For a generic weight vector (not a highest-weight vector) ${\bf \Lambda}_{W}$, one can similarly obtain $m_i -p_i$, such that
  \begin{equation}
  ( E_{-\alpha_i} )^{m_i+1} {\bf \Lambda}_{W } = ( E_{\alpha_i} )^{p_i+1} {\bf \Lambda}_{W } =0
  \end{equation}
with non-negative $m_i$ and $p_i$,   from the relation
  \begin{equation}
  m_i - p_i = \frac{2 <W , \alpha_i > }{<\alpha_i , \alpha_i >}\quad .\label{Dynkinlabelsanyweight}
\end{equation}
 \begin{figure}
\centering
\scalebox{1} 
{
\begin{pspicture}(0,-4.8498244)(10.414945,4.8573036)
\psline[linewidth=0.02cm](1.9865161,1.2114505)(8.386516,1.2114505)
\psline[linewidth=0.02cm](5.1865163,3.4114504)(5.1865163,-2.7885494)
\psline[linewidth=0.02cm,linestyle=dashed,dash=0.16cm
0.16cm](5.1865163,-4.1885495)(5.1865163,-1.9885495)
\psline[linewidth=0.02cm,linestyle=dashed,dash=0.16cm
0.16cm](5.1865163,2.6114504)(5.1865163,4.2114506)
\psline[linewidth=0.02cm,linestyle=dashed,dash=0.16cm
0.16cm](0.98651606,1.2114505)(2.7865162,1.2114505)
\psline[linewidth=0.02cm,linestyle=dashed,dash=0.16cm
0.16cm](7.786516,1.2114505)(9.586516,1.2114505)
\psline[linewidth=0.04cm,arrowsize=0.05291667cm
4.0,arrowlength=1.4,arrowinset=0.0,linecolor=red]{->}(5.1865163,1.2114505)(5.1865163,-2.1885495)
\psline[linewidth=0.04cm,arrowsize=0.05291667cm
4.0,arrowlength=1.4,arrowinset=0.0,linecolor=red]{->}(5.174273,1.2014505)(8.118759,2.9014504)
\psline[linewidth=0.04cm,arrowsize=0.05291667cm
4.0,arrowlength=1.4,arrowinset=0.0,linecolor=red]{->}(5.198759,1.2014505)(2.254273,2.9014504)
\usefont{T1}{ppl}{m}{n}
\rput{31.270283}(2.1499527,-3.0707326){\rput(6.5310473,2.3214505){$\frac{2}{3}\alpha_{1}+\frac{1}{3}\alpha_{2}$}}
\usefont{T1}{ppl}{m}{n}
\rput{90.29318}(4.4618526,-5.387741){\rput(4.8810472,-0.45854953){$-\frac{1}{3}\alpha_{1}-\frac{2}{3}\alpha_{2}$}}
\usefont{T1}{ppl}{m}{n}
\rput{-32.20001}(-0.63939524,2.3961024){\rput(3.8010473,2.3214505){$-\frac{1}{3}\alpha_{1}+\frac{1}{3}\alpha_{2}$}}
\end{pspicture}
}

\caption{\label{theweightsofthe3} The weights of the ${\bf 3}$ of $sl(3)$. All the weights have the same length and we paint them in red.}
\end{figure}
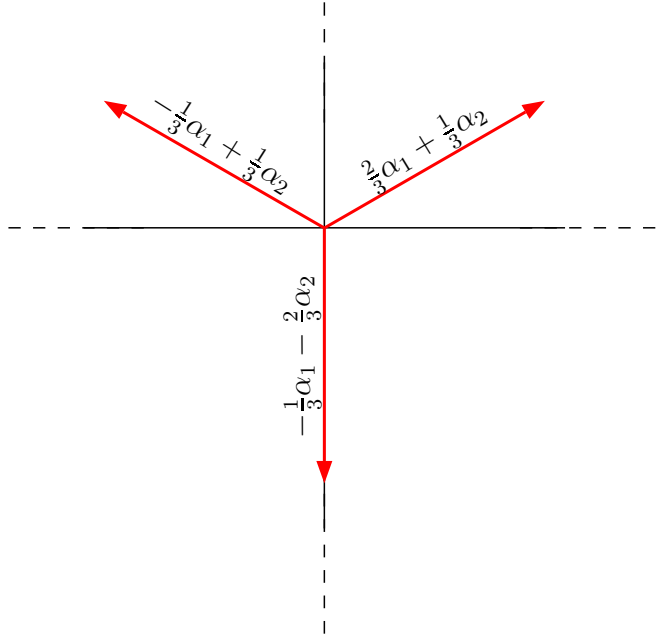
The quantities ${2 <W , \alpha_i > }/{<\alpha_i , \alpha_i >}$ are in general called {\it Dynkin labels},
 and one denotes the representation in terms of the Dynkin labels of the highest weight as $\boxed{q_1 \ q_2 \ ...\ q_r }$.
 If $q_i \neq 0$, this means that  $W_{\rm max} - \alpha_i$ is a weight, and one then obtains its Dynkin labels
by simply subtracting to each $q_j$ the following element from the the $i$-th row of the Cartan matrix $A$:
    \begin{equation}
A_{ij} = 2 \frac{<\alpha_i , \alpha_j> }{<\alpha_j , \alpha_j >}\quad .
\end{equation}
One can then read from eq. \eqref{Dynkinlabelsanyweight} which $m_j$'s are different from zero (using the fact that $p_j = \delta_{ij}$ because the weight was obtained by subtracting $\alpha_i$ to the highest weight), and correspondingly one can construct the weight $W_{\rm max} -\alpha_i -\alpha_j$, whose Dynkin labels are obtained by subtracting to the previous ones the $j$th row of the Cartan matrix. The full representation is constructed by iterating this procedure, that is by keeping subtracting simple roots. One can show that one can never act on a weight with a raising
operator in a direction different from the one the weight comes from
without annihilating it. This means that  at each stage  one knows the value of all $p_j$'s because one knows how each weight  is related to the previous ones. Thus, given the Dynkin labels of the highest weight, there is a simple iterative procedure to re-construct all the weights of the representation. The representation is complete when
one obtains a weight such that all $m_i$'s are zero. Such weight is called the {\it lowest weight}.

\begin{figure}
 \centering

\scalebox{1} 
{
\begin{pspicture}(0,-5.1296544)(11.604319,4.1296754)
\psline[linewidth=0.06cm,arrowsize=0.05291667cm
4.0,arrowlength=1.4,arrowinset=0.0,linecolor=red]{->}(5.7536855,-0.08980105)(2.9377182,1.5571133)
\psline[linewidth=0.06cm,arrowsize=0.05291667cm
4.0,arrowlength=1.4,arrowinset=0.0,linecolor=red]{->}(5.7536855,-0.08980105)(8.577701,1.5432748)
\psline[linewidth=0.02cm](8.754422,-0.08931123)(2.5544224,-0.08931123)
\psline[linewidth=0.02cm](5.743871,-4.089789)(5.754422,2.1106887)
\psline[linewidth=0.02cm,linestyle=dashed,dash=0.16cm
0.16cm](5.7679167,3.3101814)(5.762519,1.1101881)
\psline[linewidth=0.02cm,linestyle=dashed,dash=0.16cm
0.16cm](5.7448525,-3.4897902)(5.7409267,-5.0897856)
\psline[linewidth=0.02cm,linestyle=dashed,dash=0.16cm
0.16cm](9.953667,-0.10550432)(8.153672,-0.10108778)
\psline[linewidth=0.02cm,linestyle=dashed,dash=0.16cm
0.16cm](2.6344223,-0.08931123)(1.8544223,-0.08931123)
\psdots[dotsize=0.12,dotangle=179.85942](5.7536855,-0.08980105)
\usefont{T1}{ppl}{m}{n}
\rput{89.93101}(7.127371,-3.6461434){\rput(5.3589535,1.7606888){$\frac{1}{3}\alpha_{1}+\frac{2}{3}\alpha_{2}$}}
\usefont{T1}{ppl}{m}{n}
\rput{-32.132645}(0.008459474,2.4391885){\rput(4.2089534,1.2206888){$-\frac{2}{3}\alpha_{1}+\frac{2}{3}\alpha_{2}$}}
\usefont{T1}{ppl}{m}{n}
\rput{-30.994379}(1.5553126,3.8991597){\rput(7.7789536,-0.83931124){$\frac{1}{3}\alpha_{1}-\frac{1}{3}\alpha_{2}$}}
\usefont{T1}{ppl}{m}{n}
\rput{31.240772}(0.12812975,-2.0884645){\rput(3.7689536,-0.7993112){$-\frac{2}{3}\alpha_{1}-\frac{1}{3}\alpha_{2}$}}
\usefont{T1}{ppl}{m}{n}
\rput{90.00887}(3.624694,-7.134318){\rput(5.3489537,-1.7393112){$-\frac{2}{3}\alpha_{1}-\frac{4}{3}\alpha_{2}$}}
\usefont{T1}{ppl}{m}{n}
\rput{30.055645}(1.5019294,-3.4245937){\rput(7.0989537,1.1006888){$\frac{4}{3}\alpha_{1}+\frac{2}{3}\alpha_{2}$}}
\psline[linewidth=0.06cm,arrowsize=0.05291667cm
4.0,arrowlength=1.4,arrowinset=0.0,linecolor=red]{->}(5.7536855,-0.08980105)(5.7456865,-3.3497913)
\psline[linewidth=0.06cm,arrowsize=0.05291667cm
4.0,arrowlength=1.4,arrowinset=0.0]{->}(5.7536855,-0.08980105)(7.1716695,-0.91328275)
\psline[linewidth=0.06cm,arrowsize=0.05291667cm
4.0,arrowlength=1.4,arrowinset=0.0]{->}(5.7536855,-0.08980105)(4.331678,-0.90631443)
\psline[linewidth=0.06cm,arrowsize=0.05291667cm
4.0,arrowlength=1.4,arrowinset=0.0]{->}(5.7536855,-0.08980105)(5.7577095,1.550194)
\end{pspicture}
}

\caption{\label{the6ofsl3} The weights of the {\bf 6} of $sl(3)$. We have painted in red the three longest weights. }
\end{figure}
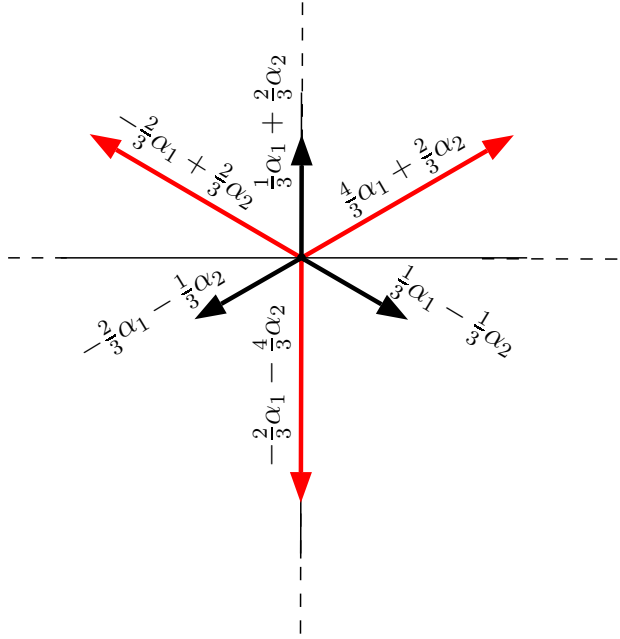

It is instructive to consider the simple example of $sl(3)$, whose Cartan matrix is
  \begin{equation}
\begin{pmatrix}
  2 & -1\\
  -1 & 2
\end{pmatrix}
\quad ,
\end{equation}
as it can be deduced from Fig.~\ref{rootssl3}.
The lowest-dimensional representation is the ${\bf 3}$, whose highest weight is denoted by the Dynkin labels $\boxed{1 \ 0}$. Writing $W_{\rm max}^{ \bf  3}$
as a linear combination of the simple roots and using eq. \eqref{qihighestweight} with $q_1=1$ and  $q_2=0$, one derives
  \begin{equation}
W_{\rm max}^{ \bf  3} = \frac{2}{3} \alpha_1 + \frac{1}{3} \alpha_2 \quad .
\end{equation}
From the fact that $q_1 =1$ and $q_2 =0$ one obtains the weight $W_{\rm max }^{\bf 3} - \alpha_1 $, whose Dynkin labels are $\boxed{ -1 \ 1}$. We know that $p_1 =1$, which implies $m_1=0$, and $p_2 =0$, which implies $m_2 =1$. We can then write the weight  $W_{\rm max }^{\bf 3} - \alpha_1  -\alpha_2$, with Dynkin labels $\boxed{0 \ -1}$.
We know that this weight has $p_1=0$ and $p_2 =1$, which imply that all $m_i$'s are zero. This is the lowest weight. All the weights of the representation are drawn in Fig.~\ref{theweightsofthe3}.
As another example, we consider the adjoint of $sl(3)$, whose highest weight has Dynkin labels $\boxed{1 \ 1 }$, which gives
  \begin{equation}
  W_{\rm max}^{\bf 8} = \alpha_1 + \alpha_2 \quad .
\end{equation}
Using the technique that we have just reviewed, one obtains all the weights of this representation, which are the roots of Fig.~\ref{rootssl3}.

In general, it can happen that the Dynkin labels are all non-negative. In such case one calls the corresponding weight a {\it dominant weight}. The highest weight is clearly a dominant weight, but the contrary is not necessarily true: there can be dominant weights that are not highest weights. We consider as an example the ${\bf 6}$ of $sl(3)$. The Dynkin labels of the highest weight are $\boxed{2 \ 0}$, corresponding to
  \begin{equation}
  W_{\rm max}^{\bf 6} = \frac{4}{3} \alpha_1 + \frac{2}{3} \alpha_2 \quad .
\end{equation}
The weights of the representation are shown in Fig.~\ref{the6ofsl3}.
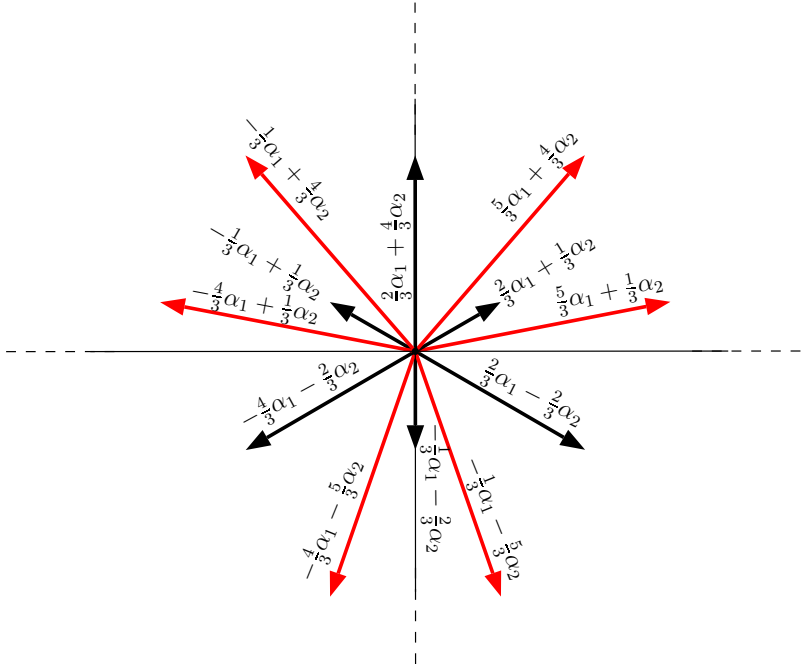
\begin{figure}
 \centering

\scalebox{.8} 
{
\begin{pspicture}(0,-6.7350965)(14.601648,5.7372804)
\psline[linewidth=0.06cm,arrowsize=0.05291667cm
4.0,arrowlength=1.4,arrowinset=0.0]{->}(7.00721,0.20367658)(7.008521,-1.4363229)
\psdots[dotsize=0.12,dotangle=180.04582](7.00721,0.20367658)
\psline[linewidth=0.06cm,arrowsize=0.05291667cm
4.0,arrowlength=1.4,arrowinset=0.0]{->}(7.00721,0.20367658)(9.82852,-1.4340677)
\psline[linewidth=0.06cm,arrowsize=0.05291667cm
4.0,arrowlength=1.4,arrowinset=0.0,linecolor=red]{->}(7.00721,0.20367658)(5.5904727,-3.8774576)
\psline[linewidth=0.06cm,arrowsize=0.05291667cm
4.0,arrowlength=1.4,arrowinset=0.0,linecolor=red]{->}(7.00721,0.20367658)(11.246552,1.0270672)
\psline[linewidth=0.06cm,arrowsize=0.05291667cm
4.0,arrowlength=1.4,arrowinset=0.0]{->}(7.00721,0.20367658)(8.426554,1.0248119)
\psline[linewidth=0.06cm,arrowsize=0.05291667cm
4.0,arrowlength=1.4,arrowinset=0.0]{->}(7.00721,0.20367658)(4.188522,-1.4385781)
\psline[linewidth=0.06cm,arrowsize=0.05291667cm
4.0,arrowlength=1.4,arrowinset=0.0,linecolor=red]{->}(7.00721,0.20367658)(9.824601,3.4659307)
\psline[linewidth=0.06cm,arrowsize=0.05291667cm
4.0,arrowlength=1.4,arrowinset=0.0]{->}(7.00721,0.20367658)(5.586554,1.0225407)
\psline[linewidth=0.06cm,arrowsize=0.05291667cm
4.0,arrowlength=1.4,arrowinset=0.0]{->}(7.00721,0.20367658)(7.0046024,3.4636755)
\usefont{T1}{ppl}{m}{n}
\rput{67.83036}(1.1150942,-6.823015){\rput(5.601501,-2.5664835){$-\frac{4}{3}\alpha_{1}-\frac{5}{3}\alpha_{2}$}}
\psline[linewidth=0.06cm,arrowsize=0.05291667cm
4.0,arrowlength=1.4,arrowinset=0.0,linecolor=red]{->}(7.00721,0.20367658)(2.766555,1.0202855)
\psline[linewidth=0.06cm,arrowsize=0.05291667cm
4.0,arrowlength=1.4,arrowinset=0.0,linecolor=red]{->}(7.00721,0.20367658)(4.184603,3.4614203)
\psline[linewidth=0.06cm,arrowsize=0.05291667cm
4.0,arrowlength=1.4,arrowinset=0.0,linecolor=red]{->}(7.00721,0.20367658)(8.430472,-3.8751864)
\usefont{T1}{ppl}{m}{n}
\rput{-30.23159}(-0.22843747,2.5097709){\rput(4.501501,1.6935165){$-\frac{1}{3}\alpha_{1}+\frac{1}{3}\alpha_{2}$}}
\usefont{T1}{ppl}{m}{n}
\rput{-26.833107}(1.2118564,3.9958003){\rput(8.951501,-0.5264835){$\frac{2}{3}\alpha_{1}-\frac{2}{3}\alpha_{2}$}}
\usefont{T1}{ppl}{m}{n}
\rput{-89.22752}(9.334733,5.256912){\rput(7.301501,-2.0864835){$-\frac{1}{3}\alpha_{1}-\frac{2}{3}\alpha_{2}$}}
\usefont{T1}{ppl}{m}{n}
\rput{31.781218}(2.2211142,-4.6066647){\rput(9.171501,1.6135166){$\frac{2}{3}\alpha_{1}+\frac{1}{3}\alpha_{2}$}}
\usefont{T1}{ppl}{m}{n}
\rput{-68.029816}(7.627771,6.0977616){\rput(8.301501,-2.5864835){$-\frac{1}{3}\alpha_{1}-\frac{5}{3}\alpha_{2}$}}
\psline[linewidth=0.02cm](12.207208,0.20783512)(1.6072114,0.19935809)
\psline[linewidth=0.02cm](7.0104084,-3.796322)(7.003851,4.403675)
\psline[linewidth=0.02cm,linestyle=dashed,dash=0.16cm 0.16cm](7.002571,6.0036745)(7.0043306,3.8036754)
\psline[linewidth=0.02cm,linestyle=dashed,dash=0.16cm 0.16cm](7.0100884,-3.3963223)(7.0113683,-4.9963217)
\psline[linewidth=0.02cm,linestyle=dashed,dash=0.16cm 0.16cm](13.4072075,0.20879479)(11.607208,0.20735529)
\psline[linewidth=0.02cm,linestyle=dashed,dash=0.16cm 0.16cm](2.0072112,0.19967797)(0.20721175,0.19823848)
\psdots[dotsize=0.12,dotangle=180.04582](7.00721,0.20367658)
\usefont{T1}{ppl}{m}{n}
\rput{-49.04286}(-0.7599882,4.80948){\rput(4.8615007,3.2535164){$-\frac{1}{3}\alpha_{1}+\frac{4}{3}\alpha_{2}$}}
\usefont{T1}{ppl}{m}{n}
\rput{90.81805}(8.754715,-4.7151318){\rput(6.671501,1.9735166){$\frac{2}{3}\alpha_{1}+\frac{4}{3}\alpha_{2}$}}
\usefont{T1}{ppl}{m}{n}
\rput{11.196235}(0.42397848,-1.9700599){\rput(10.231501,1.1935165){$\frac{5}{3}\alpha_{1}+\frac{1}{3}\alpha_{2}$}}
\usefont{T1}{ppl}{m}{n}
\rput{48.894444}(5.491604,-5.7242293){\rput(9.011501,3.1935165){$\frac{5}{3}\alpha_{1}+\frac{4}{3}\alpha_{2}$}}
\usefont{T1}{ppl}{m}{n}
\rput{-10.659738}(-0.10541875,0.8254959){\rput(4.3415008,0.99351656){$-\frac{4}{3}\alpha_{1}+\frac{1}{3}\alpha_{2}$}}
\usefont{T1}{ppl}{m}{n}
\rput{28.546108}(0.39093283,-2.501239){\rput(5.081501,-0.46648347){$-\frac{4}{3}\alpha_{1}-\frac{2}{3}\alpha_{2}$}}
\end{pspicture}
}

\caption{\label{the15ofsl3} The weights of the ${\bf 15}$ of $sl(3)$. The three shortest weights have multiplicity 2. We paint in red the 6 longest weights. }
\end{figure}
The weight $W_{\rm max}^{\bf 6} - \alpha_1 = \frac{1}{3}\alpha_1 + \frac{2}{3} \alpha_2$ has Dynkin labels $\boxed{0 \ 1}$ and thus it is a dominant weight.  If one considered this as a highest weight, one would obtain the sub-representation ${\bf \overline{3}}$ which corresponds to the black weights in the figure. The black weights are shorter than the red ones (in particular, one can notice that the difference of the squared lengths is equal to the squared length of the roots).  This result is completely general: dominant weights other than the highest weight give rise to sub-representations whose weights are shorter than the highest weight. Only representations with one dominant weight (i.e.~the highest weight) have all weights of the same length.
The case of the adjoint representation is actually a particular case of a representation with more than one dominant weight. Indeed, the Cartan generators, whose Dynkin labels are all zero, are a degenerate case of a dominant weight. As an additional example we consider the ${\bf 15}$, whose weights are shown in Fig.~\ref{the15ofsl3}. The Dynkin labels of the highest weight are $\boxed{ 2 \ 1}$, giving
 \begin{equation}
W_{\rm max}^{\bf 15} = \frac{5}{3} \alpha_1 + \frac{4}{3} \alpha_2 \quad.
\end{equation}
The dominant weight $\frac{2}{3} \alpha_1 + \frac{4}{3}\alpha_2$ has Dynkin labels $\boxed{ 0 \ 2 }$, while the dominant weight $\frac{2}{3}\alpha_1 + \frac{1}{3} \alpha_2$ (with multiplicity 2) has Dynkin labels $\boxed{1 \ 0}$. As it is clear from the figure, there are 6 long weights, 3 medium weights and 6 (3 with multiplicity 2) short weights.

\begin{figure}
 \centering
\scalebox{1} 
{
\begin{pspicture}(-1,-2.3322396)(2.281979,2.3322396)
\psline[linewidth=0.02cm](1.9767709,-0.06015625)(0.75677085,-1.8601563)
\psline[linewidth=0.02cm](0.17677084,1.8598437)(1.5367708,0.09984375)
\usefont{T1}{ppl}{m}{n}
\rput(0.4919271,-2.0251563){\large \color{white}\psframebox[linewidth=0.02,fillstyle=solid,fillcolor=red]{0 -1}}
\usefont{T1}{ppl}{m}{n}
\rput(0.40786457,1.9748437){\large \color{white}\psframebox[linewidth=0.02,fillstyle=solid,fillcolor=red]{1 0}}
\usefont{T1}{ppl}{m}{n}
\rput(1.695677,-0.02515625){\large \color{white}\psframebox[linewidth=0.02,fillstyle=solid,fillcolor=red]{-1 1}}
\end{pspicture}
}
\scalebox{1} 
{
\begin{pspicture}(0,-2.3848958)(3.2985418,2.3848958)
\psline[linewidth=0.02cm](0.76927084,2.1046875)(2.2892709,0.1446875)
\psline[linewidth=0.02cm](2.3692708,-0.0153125)(1.1492709,-1.8153125)
\usefont{T1}{ppl}{m}{n}
\rput(0.9927083,2.0196874){\large \color{white}\psframebox[linewidth=0.02,fillstyle=solid,fillcolor=red]{$T_{1}$}}
\usefont{T1}{ppl}{m}{n}
\rput(2.1927083,0.0196875){\large \color{white}\psframebox[linewidth=0.02,fillstyle=solid,fillcolor=red]{$T_{2}$}}
\usefont{T1}{ppl}{m}{n}
\rput(0.9927083,-1.9803125){\large \color{white}\psframebox[linewidth=0.02,fillstyle=solid,fillcolor=red]{$T_{3}$}}
\end{pspicture}
}
\caption{\label{3ofsl3weightsandcomponents} The Dynkin labels and the components of the ${\bf 3}$ of $sl(3)$. Note that the black lines in the left part of the Figure connect to given entries of the boxes. This indicates which root is subtracted from a box when going down the black line. In general, we paint in red all the Dynkin labels and components that are associated to the longest weights of an irreducible representation. In this case all the weights have the same length (see
Fig.~\ref{theweightsofthe3}). }
\end{figure}
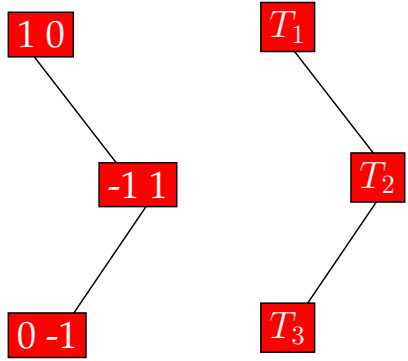

In order to determine the relation between the weights of a representation and the half-supersymmetric branes associated to the corresponding potential, it is instructive to consider the special case of $sl(n)$ algebras where there is a natural action of the creation and annihilation operators $E_{\pm \alpha}$ and of the Cartan generators $H_{\alpha}$ on the fundamental representation in terms of components. Denoting with $M$ the index of the fundamental representation, the $n-1$ generators associated to the simple roots $E_{\alpha_i}$, $i=1,...,n-1$ are the upper-triangular matrices $(T_i{}^{i+1})_M{}^N$ whose entries are 1 for $M=i$, $N=i+1$, and zero otherwise, while the Cartan generators $H_{\alpha_i}$ are diagonal matrices $( T_i{}^i )_M{}^N$ whose entries are $1/2$ for $M=N=i$, $-1/2$ for $M=N=i+1$ and zero otherwise. The annihilation operators $E_{-\alpha_i}$ are equal to $E_{\alpha_i}^\dagger$.
In $sl(n)$, summing $\alpha_i$ and $\alpha_j$ gives a root only if $i=j\pm 1$, and the root $\alpha_i + \alpha_{i+1}$ corresponds to the generator $E_{\alpha_i + \alpha_{i+1}}=[  E_{\alpha_i}, E_{\alpha_{i+1}} ]$. Realising the algebra in terms of $n\times n $ matrices as above, this leads to the matrix multiplication
\begin{equation}
 (T_i{}^{i+1})_M{}^N (T_{i+1}{}^{i+2})_N{}^P = (T_i{}^{i+2})_M{}^P \ ,
\end{equation}
which is  the upper-triangular matrix  whose entries are 1 for $M=i$, $P=i+2$, and zero otherwise. This generalises to all the positive roots: the sum of $k$ simple roots $\alpha_{i_1},\alpha_{i_2}, \alpha_{i_3},...,\alpha_{i_k}$, with $i_1 \leq i_2 \leq i_3 \leq...\leq i_k$, is a root only if $i_2 = i_1+1, i_3= i_1+2,...,i_k=i_1+k-1$, and the corresponding generator is the upper-triangular matrix $( T_{i_1}{}^{i_1 +k})_M{}^N$  whose entries are 1 for $M=i_1$, $N=i_1+k$, and zero otherwise. The whole set of positive roots thus gives all the possible real upper-triangular $n\times n$ matrices.

\begin{figure}
 \centering
\scalebox{1} 
{
\begin{pspicture}(-2,-4.3261456)(2.6788542,4.3261456)
\psline[linewidth=0.02cm](0.27364585,-2.1803124)(0.9136458,-3.9003124)
\psline[linewidth=0.02cm](2.3936458,-2.1803124)(1.5536458,-3.8603125)
\psline[linewidth=0.02cm](1.0336459,-0.1403125)(1.8936459,-1.8603125)
\psline[linewidth=0.02cm](1.4536458,-0.1003125)(0.7736458,-1.8603125)
\psline[linewidth=0.02cm](0.11364584,1.8796875)(1.0136459,0.1596875)
\psline[linewidth=0.02cm](2.4336457,1.9196875)(1.4736458,0.1796875)
\psline[linewidth=0.02cm](0.9736458,3.8396876)(1.9136459,2.1196876)
\psline[linewidth=0.02cm](1.5136459,3.8596876)(0.73364586,2.1596875)
\usefont{T1}{ppl}{m}{n}
\rput(1.1991146,3.9746876){\large \color{white}\psframebox[linewidth=0.02,fillstyle=solid,fillcolor=red]{1 1}}
\usefont{T1}{ppl}{m}{n}
\rput(1.1725521,-4.0253124){\large \color{white}\psframebox[linewidth=0.02,fillstyle=solid,fillcolor=red]{-1 -1}}
\usefont{T1}{ppl}{m}{n}
\rput(2.0925522,-2.0253124){\large \color{white}\psframebox[linewidth=0.02,fillstyle=solid,fillcolor=red]{-2 1}}
\usefont{T1}{ppl}{m}{n}
\rput(0.47739583,-2.0253124){\large \color{white}\psframebox[linewidth=0.02,fillstyle=solid,fillcolor=red]{1 -2}}
\usefont{T1}{ppl}{m}{n}
\rput(0.48723957,1.9746875){\large \color{white}\psframebox[linewidth=0.02,fillstyle=solid,fillcolor=red]{2 -1}}
\usefont{T1}{ppl}{m}{n}
\rput(2.0908334,1.9746875){\large \color{white}\psframebox[linewidth=0.02,fillstyle=solid,fillcolor=red]{-1 2}}
\usefont{T1}{ppl}{m}{n}
\rput(1.2144271,-0.0253125){\large \psframebox[linewidth=0.02,fillstyle=solid]{0 0}}
\end{pspicture}
}
\scalebox{1} 
{
\begin{pspicture}(0,-4.384896)(8.618542,4.384896)
\psline[linewidth=0.02cm](3.1292708,-2.1953125)(4.349271,-3.7553124)
\psline[linewidth=0.02cm](5.269271,-2.2553124)(4.189271,-3.8353126)
\psline[linewidth=0.02cm](4.249271,-0.2353125)(5.229271,-1.7153125)
\psline[linewidth=0.02cm](4.309271,-0.2753125)(3.1692708,-1.7353125)
\psline[linewidth=0.02cm](3.1092708,1.7646875)(4.309271,0.2646875)
\psline[linewidth=0.02cm](5.269271,1.7646875)(4.209271,0.2046875)
\psline[linewidth=0.02cm](4.269271,3.7046876)(5.249271,2.3046875)
\psline[linewidth=0.02cm](4.289271,3.6446874)(3.1092708,2.2446876)
\usefont{T1}{ppl}{m}{n}
\rput(4.1927085,4.0196877){\large \color{white}\psframebox[linewidth=0.02,fillstyle=solid,fillcolor=red]{$T_{1}^{3}$}}
\usefont{T1}{ppl}{m}{n}
\rput(3.1927083,2.0196874){\large \color{white}\psframebox[linewidth=0.02,fillstyle=solid,fillcolor=red]{$T_{1}^{2}$}}
\usefont{T1}{ppl}{m}{n}
\rput(5.1927085,2.0196874){\large \color{white}\psframebox[linewidth=0.02,fillstyle=solid,fillcolor=red]{$T_{2}^{3}$}}
\usefont{T1}{ppl}{m}{n}
\rput(4.2527084,0.0196875){\large \psframebox[linewidth=0.02,fillstyle=solid]{$T_{1}^{1},T_{2}^{2},T_{3}^{3}$}}
\usefont{T1}{ppl}{m}{n}
\rput(4.1927085,-3.9803126){\large \color{white}\psframebox[linewidth=0.02,fillstyle=solid,fillcolor=red]{$T_{3}^{1}$}}
\usefont{T1}{ppl}{m}{n}
\rput(3.1927083,-1.9803125){\large \color{white}\psframebox[linewidth=0.02,fillstyle=solid,fillcolor=red]{$T_{3}^{2}$}}
\usefont{T1}{ppl}{m}{n}
\rput(5.1927085,-1.9803125){\large \color{white}\psframebox[linewidth=0.02,fillstyle=solid,fillcolor=red]{$T_{2}^{1}$}}
\end{pspicture}
}

\caption{\label{adjofsl3weightsandcomponents} The Dynkin labels and the components of the weights of the adjoint representation of $sl(3)$. The red entries correspond to the roots (which are the longest weights of the representation). }
\end{figure}
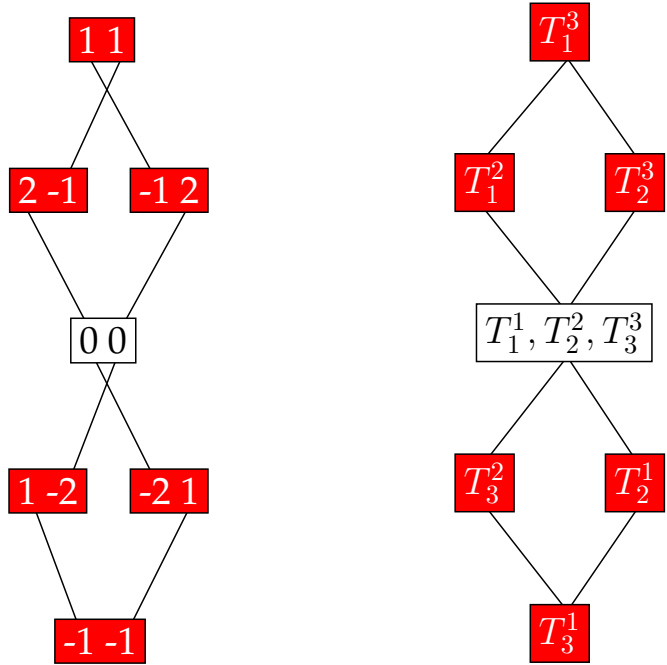

\begin{figure}
 \centering
\scalebox{1} 
{
\begin{pspicture}(-1,-4.3322396)(3.2785416,4.3322396)
\psline[linewidth=0.02cm](1.9367708,-2.1401563)(0.7967708,-3.8601563)
\psline[linewidth=0.02cm](2.956771,-0.10015625)(1.9567708,-1.9001563)
\psline[linewidth=0.02cm](0.27677083,-0.14015625)(1.5167708,-1.9401562)
\psline[linewidth=0.02cm](0.39677083,3.8798437)(1.3967708,2.1398437)
\psline[linewidth=0.02cm](1.4967709,1.8598437)(2.456771,0.13984375)
\psline[linewidth=0.02cm](1.8367709,1.8598437)(0.75677085,0.17984375)
\usefont{T1}{ppl}{m}{n}
\rput(0.49020833,-4.025156){\large \color{white}\psframebox[linewidth=0.02,fillstyle=solid,fillcolor=red]{0 -2}}
\usefont{T1}{ppl}{m}{n}
\rput(1.701302,-2.0251563){\large \psframebox[linewidth=0.02,fillstyle=solid]{-1 0}}
\usefont{T1}{ppl}{m}{n}
\rput(2.6939583,-0.02515625){\large \color{white}\psframebox[linewidth=0.02,fillstyle=solid,fillcolor=red]{-2 2}}
\usefont{T1}{ppl}{m}{n}
\rput(1.611927,1.9748437){\large \psframebox[linewidth=0.02,fillstyle=solid]{0 1}}
\usefont{T1}{ppl}{m}{n}
\rput(0.61598957,3.9748437){\large \color{white}\psframebox[linewidth=0.02,fillstyle=solid,fillcolor=red]{2 0}}
\usefont{T1}{ppl}{m}{n}
\rput(0.48223957,-0.02515625){\large \psframebox[linewidth=0.02,fillstyle=solid]{1 -1}}
\end{pspicture}
}
\scalebox{1} 
{
\begin{pspicture}(0,-4.384896)(4.5785418,4.384896)
\psline[linewidth=0.02cm](1.1692709,-0.1753125)(2.4492707,-2.1153126)
\psline[linewidth=0.02cm](2.1292708,-2.2953124)(0.9892708,-4.0153127)
\psline[linewidth=0.02cm](3.4092708,-0.0553125)(2.4092708,-1.8553125)
\psline[linewidth=0.02cm](1.1892709,3.9246874)(2.1892707,2.1846876)
\psline[linewidth=0.02cm](2.489271,1.7046875)(3.4492707,-0.0153125)
\psline[linewidth=0.02cm](2.229271,1.9046875)(1.1492709,0.2246875)
\usefont{T1}{ppl}{m}{n}
\rput(1.3327084,4.0196877){\large \color{white}\psframebox[linewidth=0.02,fillstyle=solid,fillcolor=red]{$T_{11}$}}
\usefont{T1}{ppl}{m}{n}
\rput(2.3327084,2.0196874){\large \psframebox[linewidth=0.02,fillstyle=solid]{$T_{12}$}}
\usefont{T1}{ppl}{m}{n}
\rput(1.1327083,0.0196875){\large \psframebox[linewidth=0.02,fillstyle=solid]{$T_{13}$}}
\usefont{T1}{ppl}{m}{n}
\rput(3.3327084,0.0196875){\large \color{white}\psframebox[linewidth=0.02,fillstyle=solid,fillcolor=red]{$T_{22}$}}
\usefont{T1}{ppl}{m}{n}
\rput(2.3327084,-1.9803125){\large \psframebox[linewidth=0.02,fillstyle=solid]{$T_{23}$}}
\usefont{T1}{ppl}{m}{n}
\rput(1.1327083,-3.9803126){\large \color{white}\psframebox[linewidth=0.02,fillstyle=solid,fillcolor=red]{$T_{33}$}}
\end{pspicture}
}

\caption{\label{6ofsl3weightsandcomponents} The Dynkin labels and the components of the ${\bf 6}$ of $sl(3)$. The red entries correspond to the longest weights in the representation (see Fig. \ref{the6ofsl3}).
}
\end{figure}
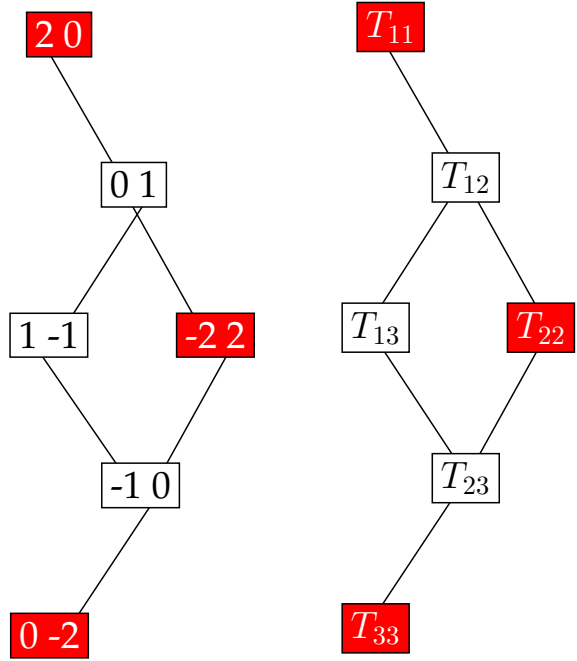

\begin{figure}
 \centering
\scalebox{1} 
{
\begin{pspicture}(-1,-6.299427)(3.1544793,6.299427)
\psline[linewidth=0.02cm](1.0483333,-2.1535938)(1.6883334,-3.8335938)
\psline[linewidth=0.02cm](1.4483334,-2.1335938)(0.66833335,-3.9135938)
\psline[linewidth=0.02cm](2.9083333,-2.0935938)(2.0683334,-3.8535938)
\psline[linewidth=0.02cm](1.3083333,1.8464062)(0.66833335,0.06640625)
\psline[linewidth=0.02cm](1.0283333,1.8664062)(1.6883334,0.10640625)
\psline[linewidth=0.02cm](2.9683332,1.9464062)(2.0683334,0.10640625)
\psline[linewidth=0.02cm](0.16833334,-0.13359375)(1.0683334,-1.8535937)
\psline[linewidth=0.02cm](2.1483333,-0.15359375)(1.4683334,-1.9335938)
\psline[linewidth=0.02cm](1.5883334,-0.13359375)(2.4083333,-1.8135937)
\psline[linewidth=0.02cm](0.38833332,3.8464062)(1.0483333,2.1064062)
\psline[linewidth=0.02cm](1.9483334,3.8664062)(1.3083333,2.0864062)
\psline[linewidth=0.02cm](1.5083333,3.8664062)(2.4283333,2.1864061)
\psline[linewidth=0.02cm](0.24833333,-4.0535936)(0.86833334,-5.893594)
\psline[linewidth=0.02cm](2.0683334,-4.1335936)(1.2283334,-5.913594)
\psline[linewidth=0.02cm](0.9683333,5.9064064)(1.8683333,4.0664062)
\psline[linewidth=0.02cm](1.3083333,5.8664064)(0.38833332,4.106406)
\usefont{T1}{ppl}{m}{n}
\rput(1.1631771,5.976406){\color{white}\psframebox[linewidth=0.02,fillstyle=solid,fillcolor=red]{2 1}}
\usefont{T1}{ppl}{m}{n}
\rput(0.43614584,-0.02359375){\psframebox[linewidth=0.02,fillstyle=solid]{-2 2}}
\usefont{T1}{ppl}{m}{n}
\rput(0.42864582,-4.023594){\color{white}\psframebox[linewidth=0.02,fillstyle=solid,fillcolor=red]{1 -3}}
\usefont{T1}{ppl}{m}{n}
\rput(1.2344271,-2.0235937){\psframebox[linewidth=0.02,fillstyle=solid]{0 -1}}
\usefont{T1}{ppl}{m}{n}
\rput(1.8375521,-0.02359375){\psframebox[linewidth=0.02,fillstyle=solid]{-1 1}}
\usefont{T1}{ppl}{m}{n}
\rput(1.1061459,-6.023594){\color{white}\psframebox[linewidth=0.02,fillstyle=solid,fillcolor=red]{-1 -2}}
\usefont{T1}{ppl}{m}{n}
\rput(1.8420833,-4.023594){\psframebox[linewidth=0.02,fillstyle=solid]{-2 0}}
\usefont{T1}{ppl}{m}{n}
\rput(2.6361458,-2.0235937){\color{white}\psframebox[linewidth=0.02,fillstyle=solid,fillcolor=red]{-3 2}}
\usefont{T1}{ppl}{m}{n}
\rput(0.6325521,3.9764063){\color{white}\psframebox[linewidth=0.02,fillstyle=solid,fillcolor=red]{3 -1}}
\usefont{T1}{ppl}{m}{n}
\rput(1.1609896,1.9764062){\psframebox[linewidth=0.02,fillstyle=solid]{1 0}}
\usefont{T1}{ppl}{m}{n}
\rput(1.7630209,3.9764063){\psframebox[linewidth=0.02,fillstyle=solid]{0 2}}
\usefont{T1}{ppl}{m}{n}
\rput(2.6397395,1.9764062){\color{white}\psframebox[linewidth=0.02,fillstyle=solid,fillcolor=red]{-2 3}}
\end{pspicture}
}
\scalebox{1} 
{
\begin{pspicture}(0,-6.245052)(4.940729,6.245052)
\psline[linewidth=0.02cm](2.0958333,-2.1942186)(1.4958333,-3.7942188)
\psline[linewidth=0.02cm](2.0958333,-2.1942186)(2.8958333,-3.7942188)
\psline[linewidth=0.02cm](3.4958334,-1.9942187)(2.6958334,-3.7942188)
\psline[linewidth=0.02cm](2.6958334,3.8057814)(3.2958333,2.4057813)
\psline[linewidth=0.02cm](3.2958333,1.8057812)(2.6958334,0.40578124)
\psline[linewidth=0.02cm](2.6958334,-0.19421875)(3.4958334,-1.7942188)
\psline[linewidth=0.02cm](1.6358334,-4.1142187)(2.2558334,-5.954219)
\psline[linewidth=0.02cm](3.0558333,-3.9942188)(2.2158334,-5.7742186)
\psline[linewidth=0.02cm](1.3558333,0.00578125)(2.2558334,-1.7142187)
\psline[linewidth=0.02cm](2.7358334,-0.01421875)(2.0558333,-1.7942188)
\psline[linewidth=0.02cm](2.0958333,1.9857812)(1.4558333,0.20578125)
\psline[linewidth=0.02cm](2.0158334,2.0057812)(2.6758332,0.24578124)
\psline[linewidth=0.02cm](1.3758334,3.9857812)(2.0358334,2.2457812)
\psline[linewidth=0.02cm](2.7358334,4.005781)(2.0958333,2.2257812)
\psline[linewidth=0.02cm](1.9558333,6.045781)(2.8558333,4.2057815)
\psline[linewidth=0.02cm](2.2958333,6.005781)(1.3758334,4.2457814)
\usefont{T1}{ppl}{m}{n}
\rput(1.6203645,-3.8842187){\color{white}\psframebox[linewidth=0.02,fillstyle=solid,fillcolor=red]{$T_{33}^{2}$}}
\usefont{T1}{ppl}{m}{n}
\rput(1.4203646,4.1157813){\color{white}\psframebox[linewidth=0.02,fillstyle=solid,fillcolor=red]{$T_{11}^{2}$}}
\usefont{T1}{ppl}{m}{n}
\rput(2.2203646,-5.8842187){\color{white}\psframebox[linewidth=0.02,fillstyle=solid,fillcolor=red]{$T_{33}^{1}$}}
\usefont{T1}{ppl}{m}{n}
\rput(1.6203645,0.11578125){\psframebox[linewidth=0.02,fillstyle=solid]{$T_{13}^{2}$}}
\usefont{T1}{ppl}{m}{n}
\rput(2.1203647,-1.8842187){\psframebox[linewidth=0.02,fillstyle=solid]{$T_{M3}^{M}$}}
\usefont{T1}{ppl}{m}{n}
\rput(2.2203646,5.915781){\color{white}\psframebox[linewidth=0.02,fillstyle=solid,fillcolor=red]{$T_{11}^{3}$}}
\usefont{T1}{ppl}{m}{n}
\rput(3.4203646,-1.8842187){\color{white}\psframebox[linewidth=0.02,fillstyle=solid,fillcolor=red]{$T_{22}^{1}$}}
\usefont{T1}{ppl}{m}{n}
\rput(3.0203645,-3.8842187){\psframebox[linewidth=0.02,fillstyle=solid]{$T_{23}^{1}$}}
\usefont{T1}{ppl}{m}{n}
\rput(2.8203645,4.1157813){\psframebox[linewidth=0.02,fillstyle=solid]{$T_{12}^{3}$}}
\usefont{T1}{ppl}{m}{n}
\rput(2.1203647,2.1157813){\psframebox[linewidth=0.02,fillstyle=solid]{$T_{M1}^{M}$}}
\usefont{T1}{ppl}{m}{n}
\rput(3.4203646,2.1157813){\color{white}\psframebox[linewidth=0.02,fillstyle=solid,fillcolor=red]{$T_{22}^{3}$}}
\usefont{T1}{ppl}{m}{n}
\rput(2.7203646,0.11578125){\psframebox[linewidth=0.02,fillstyle=solid]{$T_{M2}^{M}$}}
\end{pspicture}
}

\caption{\label{15ofsl3weightsandcomponents} The Dynkin labels and the components of the ${\bf 15}$ of $sl(3)$. The red entries correspond to the longest weights in the representation (see Fig. \ref{the15ofsl3}).}
\end{figure}
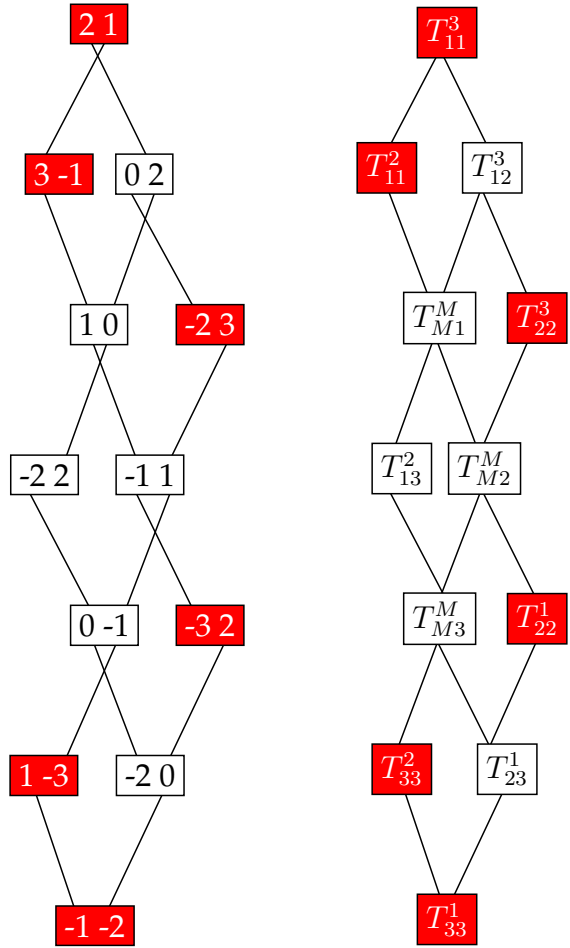

Consider again $sl(3)$ as an example. In components, the highest weight of the ${\bf 3}$ corresponds to the first component $T_1$ of a column 3-vector $T_M$. Acting with $E_{-\alpha_1}$ leads to $T_2$ and then acting with $E_{-\alpha_2}$ leads to $T_3$. This is summarised in Fig.~\ref{3ofsl3weightsandcomponents}. On the left-hand side of the figure, we write down the Dynkin labels of the weights of Fig.~\ref{theweightsofthe3}, while on the right-hand side we identify each weight with the corresponding component of $T_M$.
The same construction is given is Fig.~\ref{adjofsl3weightsandcomponents} for the case of the adjoint representation. In this case the highest weight is the root $\alpha_1 + \alpha_2$ with Dynkin labels $\boxed{1 \ 1 }$  and it corresponds to the third upper-triangular  matrix $T_1{}^3$,  which when acting on $T_1$ gives $T_3$. The Cartan generators are associated to the weight $\boxed{0 \ 0}$, which are the tensors $T_1{}^1$, $T_2{}^2$ and $T_3{}^3$ with $T_1{}^1+ T_2{}^2+ T_3{}^3 =0$, thus giving the multiplicity 2 of the weight.
 We finally consider the ${\bf 6} $ and the ${\bf 15}$ in Figs.~\ref{6ofsl3weightsandcomponents} and \ref{15ofsl3weightsandcomponents}. The ${\bf 6}$ is the symmetric product ${\bf 3} \otimes_{\rm S} {\bf 3}$, leading to the symmetric tensor $T_{MN} = T_{NM}$. The highest weight corresponds to the component $T_{11}$, and by comparing
 Figs.~\ref{6ofsl3weightsandcomponents} and \ref{the6ofsl3}
one notices that the three long weights correspond to the components $T_{11}$, $T_{22}$ and $T_{33}$, while the short weights correspond to the components $T_{12}$, $T_{13}$ and $T_{23}$. These latter components transform exactly as the components of the antisymmetric tensor $T_{[MN]}$. This antisymmetric tensor corresponds to the ${\bf \overline{3}}$, with  highest weight of Dynkin  labels $\boxed{0 \ 1}$, and this therefore explains the presence of this weight as dominant weight of the ${\bf 6}$. The ${\bf 15}$ corresponds to the irreducible tensor $T_{MN}^P = T_{NM}^P$ and satisfying $T_{1M}^1 + T_{2M}^2 + T_{3M}^3 =0$. The highest weight corresponds to the component $T_{11}^3$, and by comparing Figs. \ref{15ofsl3weightsandcomponents} and \ref{the15ofsl3}  one can notice that the six long weights correspond to the components $T_{MM}^N$, with $M\neq N$, the medium weights correspond to the components $T_{MN}^P$ with $M$, $N$ and $P$ all different and, finally, each  short weight corresponds to the components $T_{1M}^1$, $T_{2M}^2$ and $T_{3M}^3$, their sum being equal to zero, which explains the multiplicity 2 of each of these weights.  The components corresponding to the medium weights transform like $T^{PP}$, which are associated to the long weights of the representation ${\bf \overline{6}}$ whose highest weight has Dynkin labels $\boxed{0 \ 2 }$. This explains the presence of this weight as dominant weight of the ${\bf 15}$. The components corresponding to the short weights transform like the tensor $T_M$ in the ${\bf 3}$. The highest weight of this representation has Dynkin labels $\boxed{1 \ 0}$. This weight occurs as dominant weight of the ${\bf 15}$ with multiplicity 2.

This finishes our short review of the group-theoretical tools that are needed to understand the relation between branes and weights as expressed by the ``longest-weight rule'' given in the introduction. We now proceed with 
elucidating this longest-weight rule. But first 
we need to know what the actual U-duality representations of the different $(p+1)$-form potentials are. The U-duality representations of the potentials that couple to the standard branes
have been determined long ago. They follow from the representation theory of the supersymmetry algebra.
As explained in the introduction the U-duality representations of the potentials associated to all the non-standard branes of maximal supergravity have been determined over the last few years using three different techniques: closure of the supersymmetry algebra \cite{Bergshoeff:2005ac}, using properties of $E_{11}$ \cite{Riccioni:2007au,Bergshoeff:2007qi} and applying the embedding tensor technique \cite{deWit:2008ta}.

In \cite{Bergshoeff:2006gs} it was shown by an analysis of the brane effective actions that the non-standard branes of the IIB theory are fewer than the dimensions of the $\text{SL}(2,\mathbb{R})$ representations of the corresponding fields\,\footnote{From now on, we will always consider groups instead of algebras. An infinitesimal transformation of a field in a given representation under the group corresponds to the action of the generators of the algebra in that representation.}. In particular, there are two 7-branes associated to the 8-forms, that belong to the ${\bf 3}$, and two 9-branes associated to the 10-forms, that belong to the ${\bf 4}$\,\footnote{There is also an additional doublet of 10-forms in the IIB theory~\cite{Bergshoeff:2005ac}, but one cannot write down a kappa-symmetric brane effective action associated to it.
This is in accordance with the longest-weight rule.}. This analysis was generalised to any maximal supergravity theory in any dimensions~\cite{Bergshoeff:2010xc,Bergshoeff:2011qk,Bergshoeff:2012ex}, revealing
that when one considers the potentials that couple to the non-standard branes, one always
finds that the number of branes is less than the dimension of the U-duality representation. This
is in sharp contrast with the case of standard branes, where the number of half-supersymmetric branes is always the same as the dimension of the corresponding U-duality representation. Below we will show in a few explicit examples that
this corresponds to  the fact that while the representations of the standard branes only contain one dominant weight (the highest weight), those of the non-standard branes always contain more than one dominant weight. In the latter case the branes correspond to the longest weights in the representation (those with the same length as the highest weight).

To show how this works it is instructive to consider an explicit example, that is the non-standard branes of eight-dimensional maximal supergravity, whose global symmetry is $\text{SL}(3,\mathbb{R}) \times \text{SL}(2,\mathbb{R})$.
There are 6 defect branes in the adjoint of $\text{SL}(3,\mathbb{R})$. Their corresponding  6-form potential is $A_{6,M}{}^N$, which is contracted in the Wess-Zumino term by the 5-brane charge $T^M{}_N$ with $M\neq N$ \cite{Bergshoeff:2011se}. As we have shown, these directions correspond to the roots of $\text{SL}(3,\mathbb{R})$, which clearly are the longest weights of the representation (see Figs. \ref{rootssl3} and  \ref{adjofsl3weightsandcomponents}). The 6-brane charges of  the domain walls are $T_{MNa}$ $(a=1,2)$ in the $({\bf 6,2})$. There are 6 domain walls, corresponding to the charges $T_{11a}$, $T_{22a}$ and $T_{33a}$ \cite{Bergshoeff:2012pm}. Looking at Figs. \ref{6ofsl3weightsandcomponents} and \ref{the6ofsl3}, we see  that these components correspond to the longest weights of the ${\bf 6}$ of $\text{SL}(3,\mathbb{R})$. Finally, there are six half-supersymmetric space-filling branes with 7-brane charges $T_{MN}{}^P$ in the $({\bf 15,1})$ such that $M=N$ and $P\neq M$. From Figs.~\ref{15ofsl3weightsandcomponents} and \ref{the15ofsl3} we know that these components precisely correspond to the longest weights of the representation.

\begin{figure}[h]
\begin{center}
\begin{picture}(220,70)
\multiput(10,10)(40,0){2}{\circle{10}}
\multiput(130,10)(40,0){3}{\circle{10}}
\put(15,10){\line(1,0){30}}
\put(55,10){\line(1,0){5}}
\put(67,10){\line(1,0){10}}
\put(85,10){\line(1,0){10}}
\put(103,10){\line(1,0){10}}
\put(120,10){\line(1,0){5}}
\multiput(135,10)(40,0){2}{\line(1,0){30}} \put(130,50){\circle{10}}
\put(130,15){\line(0,1){30}} \put(8,-8){$1$}
\put(48,-8){$2$} \put(117,-8){$n-3$}
\put(157,-8){$n-2$} \put(197,-8){$n-1$} \put(140,47){$n$}
\end{picture}
\caption{\sl The Dynkin diagrams of $E_6$, $E_7$  and $E_8$. \label{Ed+1diagram}}
\end{center}
\end{figure}
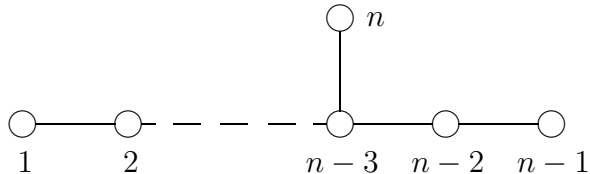

We find that the above result is completely general. The defect branes in any dimension always correspond to the components of the adjoint representation associated to the roots. Given that the symmetry groups of maximal supergravities are all simply laced, all the roots have the same length, and thus the number of defect branes is always ${\rm dim}\,G - {\rm rank}\,G$, where the Cartan generators are $\boxed{0\ 0\ ...\ 0}$ dominant weights with ${\rm rank}\,G$ multiplicity. Similarly, for all domain walls and space-filling branes one can determine all the dominant weights of the associated representations, and the number of weights that have the same length as each dominant weight. Counting the weights of the same length as the highest weight reproduces the number of half-supersymmetric branes determined in
\cite{Bergshoeff:2010xc,Bergshoeff:2011qk,Bergshoeff:2012ex,Kleinschmidt:2011vu}. The result is summarised in Table \ref{dominantweightsofnonstandardbranes}. For the exceptional cases $\text{E}_{6(6)}$, $\text{E}_{7(7)}$ and $\text{E}_{8(8)}$ we label the Dynkin weights following the numbering of the nodes of the Dynkin diagrams of
Fig.~\ref{Ed+1diagram}.

This finishes our discussion of the relation between branes and weights. In the next section we will show that the property of the representations of non-standard branes of having more than one dominant weight naturally leads to a second difference with the standard branes, namely a degeneracy of the BPS conditions.

\begin{table}\small
\begin{center}
\renewcommand{\arraystretch}{1.15}
\scalebox{0.83}{
\begin{tabular}{|c|c|c|c|c|c|c|}
\hline \rule[-1mm]{0mm}{1mm} $D$ & $G$ & repr. & codim. &  dominant weights & weights of same length\\
\hline \hline \rule[-1mm]{0mm}{1mm}  8 & $\text{SL}(3,\mathbb{R})$  & $({\bf 6,2})$ & 1 &  ${\color{red}\boxed{2 \ 0} \times \boxed{1} }$ & ${\color{red}3 \times 2}$ \\
 \rule[-1mm]{0mm}{1mm} & $\times$ &  & & $\boxed{0 \ 1} \times \boxed{1} $ & $3 \times 2 $\\
\cline{3-6} \rule[-1mm]{0mm}{1mm} &$ \text{SL}(2,\mathbb{R})$& $({\bf 15,1})$ & 0  &  ${\color{red}\boxed{2 \ 1} \times \boxed{0}} $ & ${\color{red}6} $\\
 \rule[-1mm]{0mm}{1mm}  & & & & $\boxed{0 \ 2} \times \boxed{0} $ & $3  $\\
 \rule[-1mm]{0mm}{1mm}  & & & & $2 \times \boxed{1 \ 0} \times \boxed{0} $ & $2 \times 3  $\\
\hline  \rule[-1mm]{0mm}{1mm} 7 & $\text{SL}(5,\mathbb{R})$ &${\bf \overline{40}}$ & 1 &  ${\color{red} \boxed{1 \ 1\ 0  \ 0}} $ & ${\color{red} 20}  $\\
 \rule[-1mm]{0mm}{1mm} & & & & $2 \times \boxed{0 \ 0 \ 1\ 0}  $ & $2 \times 10  $\\
\cline{3-6}
 \rule[-1mm]{0mm}{1mm}  & &${\bf \overline{15}}$ & $1_T$ &  ${\color{red} \boxed{0\ 0\ 0 \ 2}} $ & ${\color{red} 5}  $\\
 \rule[-1mm]{0mm}{1mm} & &  & & $\boxed{0 \ 0 \ 1\ 0}  $ & $10  $\\
\cline{3-6}
 \rule[-1mm]{0mm}{1mm}  & &${\bf {70}}$ & 0 & ${\color{red} \boxed{2\ 0\ 0 \ 1}} $ & ${\color{red} 20}  $\\
 \rule[-1mm]{0mm}{1mm} & & &  & $\boxed{0 \ 1 \ 0\ 1}  $ & $30  $\\
 \rule[-1mm]{0mm}{1mm} & & &  & $4 \times \boxed{1 \  0 \ 0 \  0}  $ & $4 \times 5  $\\
\hline  \rule[-1mm]{0mm}{1mm} 6 & $\text{SO}(5,5)$ &${\bf {144}}$ & 1 & ${\color{red} \boxed{1 \ 0 \ 0\ 0\ 1  }} $ & ${\color{red} 80}  $\\
 \rule[-1mm]{0mm}{1mm} & & &  & $4  \times \boxed{0 \ 0 \ 0\ 1\ 0 }  $ & $4 \times 16  $\\
\cline{3-6}
 \rule[-1mm]{0mm}{1mm}  & &${\bf {320}}$ & 0 &  ${\color{red} \boxed{1 \ 1\ 0 \ 0  \ 0}} $ & ${\color{red} 80}  $\\
 \rule[-1mm]{0mm}{1mm} & & & &  $2 \times \boxed{0 \ 0 \ 1 \ 0\ 0}  $ & $2 \times 80  $\\
 \rule[-1mm]{0mm}{1mm} & & & &  $8 \times \boxed{1 \ 0  \ 0 \ 0\ 0}  $ & $8 \times 10  $\\
\cline{3-6}
 \rule[-1mm]{0mm}{1mm}  & &${\bf {\overline{126}}}$ & $0_T$ &  ${\color{red} \boxed{0\ 0\ 0 \ 0 \ 2}} $ & ${\color{red} 16}  $\\
 \rule[-1mm]{0mm}{1mm} & &  & & $\boxed{0 \ 0 \ 1\ 0 \ 0}  $ & $80  $\\
 \rule[-1mm]{0mm}{1mm} & & & &  $3 \times \boxed{1 \  0 \ 0 \ 0 \  0}  $ & $3  \times 10  $\\
\hline  \rule[-1mm]{0mm}{1mm} 5 & $\text{E}_{6(6)}$ &${\bf \overline{351}}$ & 1  & ${\color{red} \boxed{0 \ 0 \ 0\ 1\ 0 \ 0  }} $ & ${\color{red} 216}  $\\
 \rule[-1mm]{0mm}{1mm} & & &  & $5  \times \boxed{1 \ 0 \ 0\ 0\ 0 \ 0}  $ & $5 \times 27  $\\
\cline{3-6}
 \rule[-1mm]{0mm}{1mm}  & &${\bf \overline{1728}}$ & 0 & ${\color{red} \boxed{0 \ 0\ 0 \ 0 \ 1 \ 1  }} $ & ${\color{red} 432}  $\\
 \rule[-1mm]{0mm}{1mm} & & & & $4 \times \boxed{0 \ 1 \ 0 \ 0\ 0\ 0}  $ & $4 \times 216  $\\
 \rule[-1mm]{0mm}{1mm} & & &  & $16 \times \boxed{0\ 0 \ 0  \ 0 \ 1\ 0}  $ & $16 \times 27  $\\
\hline  \rule[-1mm]{0mm}{1mm} 4 & $\text{E}_{7(7)}$ &${\bf {912}}$ & 1  & ${\color{red} \boxed{0\ 0 \ 0 \ 0\ 0 \ 0  \ 1  }} $ & ${\color{red} 576}  $\\
 \rule[-1mm]{0mm}{1mm} & & & &  $6  \times \boxed{1 \ 0 \ 0\ 0\ 0 \ 0}  $ & $6 \times 56  $\\
\cline{3-6}
 \rule[-1mm]{0mm}{1mm}  & &${\bf {8645}}$ & 0 & ${\color{red} \boxed{0 \ 0\ 0 \ 0 \ 1 \ 0\ 0  }} $ & ${\color{red} 2016}  $\\
 \rule[-1mm]{0mm}{1mm} & & &  & $5 \times \boxed{0 \ 1 \ 0 \ 0\ 0\ 0\ 0}  $ & $5 \times 756  $\\
 \rule[-1mm]{0mm}{1mm} & & &  & $22 \times \boxed{0\ 0\ 0 \ 0  \ 0 \ 1\ 0}  $ & $22 \times 126  $\\
 \rule[-1mm]{0mm}{1mm} & & &  & $77 \times \boxed{0\ 0\ 0 \ 0  \ 0 \ 0 \ 0}  $ & $77 \times 1  $\\

\hline  \rule[-1mm]{0mm}{1mm} 3 & $\text{E}_{8(8)}$ &${\bf {3875}}$ & 1  & ${\color{red} \boxed{0\ 0 \ 0 \ 0\ 0 \ 0  \ 1 \ 0 }} $ & ${\color{red} 2160}  $\\
 \rule[-1mm]{0mm}{1mm} &  & & & $7  \times \boxed{1 \ 0 \ 0\ 0\ 0 \ 0\ 0}  $ & $7 \times 240  $\\
 \rule[-1mm]{0mm}{1mm} &  & & & $35 \times \boxed{0\ 0\ 0 \ 0  \ 0 \ 0 \ 0\ 0}  $ & $35 \times 1  $\\
\cline{3-6}
 \rule[-1mm]{0mm}{1mm}  & &${\bf {147250}}$ & 0 & ${\color{red} \boxed{0 \ 0\ 0 \ 0 \ 0 \ 0\ 0\ 1  }} $ & ${\color{red} 17280}  $\\
 \rule[-1mm]{0mm}{1mm} & & &  & $6 \times \boxed{0 \ 1 \ 0 \ 0\ 0\ 0\ 0\ 0}  $ & $6 \times 6720  $\\
 \rule[-1mm]{0mm}{1mm} & & &  & $29 \times \boxed{0\ 0\ 0\ 0 \ 0  \ 0 \ 1\ 0}  $ & $29 \times 2160  $\\
 \rule[-1mm]{0mm}{1mm} & & & &$111 \times \boxed{1\ 0\ 0\ 0 \ 0  \ 0 \ 0 \ 0}  $ & $111 \times 240  $\\
 \rule[-1mm]{0mm}{1mm} & & &  & $370 \times \boxed{0\ 0\ 0 \ 0  \ 0 \ 0 \ 0\ 0}  $ & $370 \times 1  $\\
\hline
\end{tabular}
}
\caption{ \label{dominantweightsofnonstandardbranes}
The potentials associated to the $p$-branes of codimension 1 (domain walls) and codimension 0 (space-filling branes). For $p\neq 5$, there is always a single irreducible representation, and the corresponding brane supports a vector multiplet. For $p =5$, there are two representations, one corresponding to a vector brane and one to a tensor brane, and the tensor brane is identified by the subscript $T$ in the codimension entry.  In all cases, the branes correspond to the longest weights, that is the weights of the same length as the highest weight, for each representation. Their number, as well as the Dynkin labels of the highest weight, is painted in red.
}
\end{center}
\end{table}

\section{Central charges and degeneracies}

In the previous section we have given a group-theoretic characterisation of the difference between standard and non-standard branes. We have seen that the potentials corresponding to standard branes belong to representations of the global symmetry group with only one dominant weight, while the potentials corresponding to non-standard branes belong to representations with more than one dominant weight. In all cases the branes are associated to the weights within the representation that have the same length as the highest weight. For standard branes, the representations have weights which are all of the same length and therefore the number of branes is the same as the dimension of the representation, while for non-standard branes there are sets of weights of different length, each corresponding to a different dominant weight, and the half-supersymmetric branes correspond to the weights of maximum length, which is the length of the highest weight. This implies that the number of non-standard branes is less than the dimension of the representation.
 In \cite{Bergshoeff:2011se,Bergshoeff:2012pm,Bergshoeff:2012jb} it was shown that there is another crucial difference between standard and non-standard branes: while for standard branes there is a one-to-one relation between half-supersymmetric branes and BPS conditions, in the case of non-standard branes this relation is many-to-one, i.e.~more branes give rise to the same BPS condition. This has the important consequence that one can consider half-BPS configurations corresponding to bound states of different half-BPS branes which correspond to the same BPS condition. Using group-theory arguments we will show in this section  why non-standard branes can have degenerate BPS conditions.

The number of different BPS conditions that can be imposed on a half-supersymmetric $p$-brane is equal to the number of central charges of rank $p$. In maximal supersymmetric theories, the central charges form representations of the R-symmetry $H$, which is the maximal compact subgroup of the maximally non-compact U-duality group $G$. The representations of the central charges of various rank in any dimension are given in Table \ref{centralchargetable}. The table only contains central charges of rank up to $[D/2]$, because the charges of rank $p > [D/2]$ are equal to the charges of rank $D-p$ by Hodge duality. This means that for instance in $D=7$ the $p=4$ charges (associated to defect branes)  are in the ${\bf 10}$, the $p=5$ charges (associated to domain walls) are in the ${\bf 1+ 5}$ and the $p=6$ charges (associated to space-filling branes) are in the ${\bf 5}$\,\footnote{For $p=1$, only the charges other than the momentum operator can be dualised giving a $D-1$ charge for space-filling branes.}.

\begin{table}[t]
\begin{center}
\begin{tabular}{|c|c|c|c|c|c|c|c|}
\hline
$D$&$H$&$p=0$&$p=1$&$p=2$&$p=3$&$p=4$&$p=5$\\[.1truecm]
\hline \hline \rule[-1mm]{0mm}{6mm} IIA&${\unity}$&{\bf 1}&{\bf 1} +{\bf 1}&{\bf 1}&--&{\bf 1}&${\bf 1}$\\[.05truecm]
\hline \rule[-1mm]{0mm}{6mm} IIB&SO(2)&--&{\bf 1} +{\bf 2}&--&{\bf 1}&--&${\bf 1}^+ + {\bf 2}^+$\\[.05truecm]
\hline \rule[-1mm]{0mm}{6mm} 9&SO(2)&${\bf 1}+{\bf 2}$&{\bf 1}+{\bf 2}&{\bf 1}&{\bf 1}&${\bf 1}+{\bf 2}$&\\[.05truecm]
\hline \rule[-1mm]{0mm}{6mm} 8&U(2)& $ 2 \times {\bf 3}$ &{\bf 1}+{\bf 3} &
$2 \times {\bf 1}$& ${\bf 1} + {\bf 3}$
&${\bf 3}^+ + {\bf 3}^-$&\\[.05truecm]
\hline \rule[-1mm]{0mm}{6mm} 7&Sp(4)& {\bf 10} &{\bf 1}+ {\bf 5} & ${\bf 1} +{\bf 5}$ &{\bf 10} &&\\[.05truecm]
\hline \rule[-1mm]{0mm}{6mm} 6&Sp(4)$\times$Sp(4)&$({\bf 4},{\bf 4})$&$({\bf 1},{\bf 1})+({\bf 1},{\bf 1})$ & $({\bf 4},{\bf 4})$& $({\bf 10},{\bf 1})^+$ &&\\[.05truecm]
\rule[-1mm]{0mm}{6mm} & & &$+({\bf 1},
{\bf 5})+({\bf 5},
{\bf 1})$ & &$+({\bf 1},{\bf 10})^-$ &&\\[.05truecm]
\hline \rule[-1mm]{0mm}{6mm} 5&Sp(8)& {\bf 1} + {\bf 27}&{\bf 1}+ {\bf 27}& {\bf 36}&&&\\[.05truecm]
\hline \rule[-1mm]{0mm}{6mm} 4&SU(8)& ${\bf 28} + {\bf \overline{28}}$ &{\bf 1}+ {\bf 63} & ${\bf 36}^+ + {\bf \overline{36}}^-$ &&&\\[.05truecm]
\hline \rule[-1mm]{0mm}{6mm} 3&SO(16)& {\bf 120} & {\bf 1}+{\bf 135} &&&&\\[.05truecm]
\hline
\end{tabular}
\end{center}
  \caption{\sl This table indicates the representations of the  R-symmetry $H$ of the
  $p$-form central charges of $3\le D\le 10$ maximal supergravity.
  If applicable, we have also indicated the space-time duality of
  the central charges with a superscript $\pm$. There is always a singlet $p=1$ charge which is the momentum operator. }
  \label{centralchargetable}
\end{table}

In order to determine the half-BPS branes, one has to decompose the representation of $G$ of the brane charges $T$ in representations of $H$. For standard branes, the representations of $H$ one obtains are all contained in Table \ref{centralchargetable} for any $p$ and in any dimension. This means that for each component of the representation of
$G$ of a $p$-brane charge $T$ in $D$ dimensions, there is a rank $p$ central charge $Q$ in the supersymmetry algebra. For non-standard branes the situation is different: in this case the decomposition contains the associated central charge $Q$ of the correct rank, but it also contains additional representations that are not contained in the Table. We denote
these additional representations collectively by $R$. Summarising, one has schematically
  \begin{eqnarray}
  & & {\rm standard \ branes}: \ \hskip .75truecm  T \rightarrow Q \,,\nonumber \\[.2truecm]
  & & {\rm non}\text{-}{\rm standard \ branes}: \ T \rightarrow R + Q \quad .\label{standardTQnonstandardTRQ}
  \end{eqnarray}

As an example we  consider the non-standard branes in $D=7$. The 4-brane charges are in the ${\bf 24}$ (adjoint) of $\text{SL}(5,\mathbb{R})$, which decomposes under the R-symmetry $\text{Sp}(4)$ as
 \begin{equation}
 {\bf 24 } \rightarrow {\bf 14} + {\bf 10} \quad .
\end{equation}
From Table \ref{centralchargetable} one notices that only the representation ${\bf 10}$ of $\text{SO}(5)$ is present as a $p=4$ (i.e.~dual $p=3$) central charge. The fact that only part of the representation goes to the representation of the central charge explains the fact that there is a degeneracy of the BPS conditions. To understand this, it is
 instructive to analyse the decomposition in terms of components. The ${\bf 24}$ of $\text{SL}(5,\mathbb{R})$ corresponds to a traceless tensor $T_M{}^N$, with $M,N =1,...,5$. This decomposes under $\text{Sp}(4)$ in a symmetric traceless tensor $R_{MN}$ in the ${\bf 14}$ and an antisymmetric tensor $Q_{MN}$ in the ${\bf 10}$, where now $M,N$ are indices in the ${\bf 5}$ of $\text{Sp}(4)$.  The diagonal  components of the ${\bf 24}$  (i.e.~the 4 directions along the Cartan generators), that are entirely contained in the ${\bf 14}$, are not associated to any central charge and these are precisely
  the components that do not correspond to half-supersymmetric branes. The other 10 components correspond to the same central charges as the ${\bf 10}$ and this
 is the reason for the degeneracy 2.
In terms of components the degeneracy arises as follows: the components $T_M{}^N$ and $T_N{}^M$ decompose as $R_{MN}+ Q_{MN}$ and $R_{MN} - Q_{MN}$, corresponding  to the same central charge $Q_{MN}$ (up to a sign).

As another example we  consider the $D=7$ tensor 5-branes which have  brane charges $T^{MN}=T^{NM}$  in the ${\bf \overline{15}}$. This decomposes under $\text{Sp}(4)$ as
  \begin{equation}
{\bf \overline{15}} \rightarrow {\bf 14 } + {\bf 1} \quad .
\end{equation}
One can see from Table \ref{centralchargetable} that only the singlet corresponds to a $p=5$ (i.e. $p=2$) central charge in $D=7$. In components this means that a symmetric tensor $T^{MN}$ of $\text{SL}(5,\mathbb{R})$ decomposes in a symmetric traceless tensor $R_{MN}$ and the  trace part $\delta_{MN} Q$ under $\text{Sp}(4)$. All the components of $T^{MN}$ that are not diagonal have no projection on the singlet and thus are not associated to branes, while all the five diagonal components, that regardless of $R_{MN}$ have the same projection on the singlet, give rise to the same BPS condition with degeneracy 5.

The same analysis applies to the remaining non-standard branes in $D=7$. The vector domain walls correspond to 5-brane charges in the ${\bf \overline{40}}$, which decomposes as
  \begin{equation}
  {\bf \overline{40}} \rightarrow {\bf 35} + {\bf 5} \quad .
\end{equation}
One can see that only the ${\bf 5}$ representation is present in Table \ref{centralchargetable} as a $p=5$
(i.e.~dual $p=2$) central charge in $D=7$. In components, the charge $T_{MN,P}$, antisymmetric in $MN$ and such that $T_{[MN,P]}=0$, has a non-zero projection on the ${\bf 5}$ only if $P=M$ or $P=N$, in which case the representation decomposes to $Q_{[M} \delta_{N]P}$ (ignoring the part  along the ${\bf 35}$). This implies that 4 different choices of $N$ lead
 to the same charge $Q_M$, for fixed $M$. Therefore, each charge $Q_M$  has degeneracy 4. Finally, the space-filling branes correspond to the 6-brane charges $T_{MN}^P$ in the ${\bf 70}$ (which is symmetric in $MN$ and such that $T_{MN}^N =0$), which decomposes according to
  \begin{equation}
  {\bf 70} \rightarrow {\bf 35} + {\bf 30} + {\bf 5} \quad .
\end{equation}
We see that only the ${\bf 5}$ representation appears in Table \ref{centralchargetable} as a $p=6$ (i.e.~dual $p=1$) central charge. The projection of the ${\bf 70}$ on the ${\bf 5}$ of $\text{Sp}(4)$ is
$T_{MN}^P \rightarrow \delta_{MN} Q^P$, and
the components that have non-zero projection on the ${\bf 5}$ are the 20 components $T_{MM}^P$, with $M\neq P$, which implies a degeneracy 4 for the central charge.

\begin{table}\small
\begin{center}
\begin{tabular}{|c|c|c|c|c|c|c|}
\hline \rule[-2mm]{0mm}{1mm} $D$ & $H$ & repr. of $G$ & decomposition under $H$ & degeneracy& \# of branes & codim.\\
\hline \hline \rule[-2mm]{0mm}{1mm}  8 & $\text{U}(2) $ & $({\bf 8,1})$ & ${\bf 5}_0 +{\color{red}{\bf 3}_0 }$ & 2 &6 & 2\\
\cline{3-7}
 \rule[-2mm]{0mm}{1mm} & & $({\bf 1,3})$& ${\bf 1}_{+2} + {\bf 1}_{-2} + {\color{red} {\bf 1}_0} $ & 2&2 & $2_T$\\
\cline{3-7} \rule[-2mm]{0mm}{1mm} & &  $({\bf 6,2})$&  ${\bf 5}_{+1} +{\bf 5}_{-1} +
{\color{red}{\bf 1}_{+1} + {\bf 1}_{-1}} $ & 3 & 6 & 1 \\
\cline{3-7}
 \rule[-2mm]{0mm}{1mm} & & $({\bf 15,1})$& ${\bf 7}_0 + {\bf 5}_0 + {\color{red}{\bf 3}_0} $ & $2  $& 6 & 0\\
\hline
\hline  \rule[-2mm]{0mm}{1mm} 7 & $\text{Sp}(4)$ &${\bf {24}}$ & ${\bf 14} +{\color{red} \bf{10}} $ & $2  $& 20 & 2\\
\cline{3-7}
\rule[-2mm]{0mm}{1mm}  & &${\bf \overline{40}}$ & ${\bf 35} + {\color{red} \bf{5}} $ & $4  $& 20 & 1\\
\cline{3-7}
 \rule[-2mm]{0mm}{1mm}  & &${\bf \overline{15}}$ & ${\bf 14} +{\color{red} \bf 1} $ & $5  $& 5 & $1_T$\\
\cline{3-7}
 \rule[-2mm]{0mm}{1mm}  & &${\bf {70}}$ & ${\bf 35} + {\bf 30} +{\color{red} \bf 5} $ & $4 $& 20 & 0\\
\hline \hline  \rule[-2mm]{0mm}{1mm} 6 & $\text{Sp}(4)$ &${\bf {45}}$ & $({\bf 5,5}) + {\color{red}({\bf 10,1})+ ({\bf 1,10})}
 $ & 2 & 40 & 2\\
\cline{3-7}
 \rule[-2mm]{0mm}{1mm} & $ \times $&${\bf {144}}$ & $({\bf 16,4})+ ({\bf 4,16}) + {\color{red}({\bf 4,4})} $ & $5  $ & 80 &1\\
\cline{3-7}
 \rule[-2mm]{0mm}{1mm}  & $\text{Sp}(4)$ &${\bf {320}}$ & $({\bf 14,5})+ ({\bf 5,14}) +({\bf 35,1})+ ({\bf 1,35}) $ & 8 & 80 & 0
\\
 \rule[-2mm]{0mm}{1mm} & & & $+ ({\bf 5,10})+({\bf 10,5}) + {\color{red}({\bf 5,1})+({\bf 1,5}) } $ & & &  \\
\cline{3-7}
 \rule[-2mm]{0mm}{1mm}  & &${\bf {\overline{126}}}$ & $({\bf 10,10}) + ({\bf 5,5}) + {\color{red}({\bf 1,1})}$ & 16 & 16 & $0_T$\\
\hline \hline  \rule[-2mm]{0mm}{1mm} 5 & $\text{Sp}(8)$ &${\bf {78}}$ & ${\bf 42} + {\color{red}\bf 36}$  & 2 & 72 & 2\\
\cline{3-7}
 \rule[-2mm]{0mm}{1mm} & &${\bf {351}}$ & $ {\bf 315} + {\color{red}{\bf 36}}$ & 6& 216 & 1 \\
\cline{3-7}
 \rule[-2mm]{0mm}{1mm}  & &${\bf \overline{1728}}$ & ${\bf 792} +{\bf 594}+{\bf 315}+{\color{red} \bf 27}$ & 16 & 432 & 0\\
\hline  \hline \rule[-2mm]{0mm}{1mm} 4 & $\text{SU}(8)$ &${\bf {133}}$ & $ {\bf 70}+{\color{red} \bf 63}$ & 2 & 126 & 2 \\
\cline{3-7}
 \rule[-2mm]{0mm}{1mm}  & &${\bf {912}}$ & ${\bf 420} + {\bf \overline{420}}+ {\color{red}{\bf 36}+{\bf \overline{36}}}$  & 8  & 576 & 1 \\
\cline{3-7}
 \rule[-2mm]{0mm}{1mm}  & &${\bf {8645}}$ &  $ {\bf 3584} + {\bf 2352}+{\bf 945}+{\bf \overline{945}}$ & 32 & 2016 & 0\\
 \rule[-2mm]{0mm}{1mm}  & & &  $+ {\bf 378}+{\bf \overline{378}}+ {\color{red}{\bf 63}}$ & &  & \\
\hline \hline  \rule[-2mm]{0mm}{1mm} 3 & $\text{SO}(16)$ &${\bf {248}}$ & ${\bf 128} +{\color{red}{\bf 120}}$ & 2 &240 & 2\\
\cline{3-7}
 \rule[-2mm]{0mm}{1mm} &  &${\bf {3875}}$ & ${\bf 1820} +{\bf \overline{1920}}+{\color{red}{\bf 135}}$  & $16  $ & 2160 & 1\\
\cline{3-7}
 \rule[-2mm]{0mm}{1mm}  & &${\bf {147250}}$ & ${\bf 60060}+{\bf \overline{56320}}+{\bf 15360} $ & 128 & 17280 & 0\\
 \rule[-2mm]{0mm}{1mm}  & &  & $+{\bf 7020}+{\bf \overline{6435}}+{\bf \overline{1920}}+ {\color{red}{\bf 135}} $ &  & &  \\
\hline
\end{tabular}
\caption{ \label{nonstandardcentralcharge} The decomposition of the representations of the non-standard branes with respect to the R-symmetry $H$ in any dimension. In each case, the representation of the central charge is painted in red. The dimension of this representation times the degeneracy gives the number of branes. In the last column we specify the codimension of the brane, and we introduce a subscript $T$ for the tensor branes (see the caption of Table \ref{dominantweightsofnonstandardbranes}).
}
\end{center}
\end{table}

The result we find for the non-standard branes in $D=7$ is completely general. In any dimension the representations of the non-standard branes decompose under $H$ into the representation of the corresponding central charge plus additional representations. Moreover, only the components corresponding to half-supersymmetric branes (that as we know from the previous section are those associated to the longest weights) have non-zero projection on the representation of $H$ corresponding to the central charge. This projection occurs with a given degeneracy: a given central charge component corresponds to more branes. This gives the degeneracy of the BPS conditions. The general result is summarised in Table \ref{nonstandardcentralcharge}, where for any representation associated to non-standard branes in $D\leq 8$ we give the decomposition under $H$ and the multiplicity of the BPS conditions.

\section{Orbits and Invariants}

In this section we wish to consider the orbits of the half-supersymmetric branes under infinitesimal
U-duality transformations. The orbits of the standard branes of maximal supergravity theories have been derived long ago in \cite{Lu:1997bg,Ferrara:1997uz}.
Here we will consider the orbits of the non-standard branes as well and point out what the differences with the orbits of the
standard branes are.

In general,  under the algebra $g$, a weight can either  transform infinitesimally to the other weights or stay invariant. The generators that leave the weight invariant form a subalgebra of $g$ which is the stabiliser of the weight orbit. Therefore, all the half-supersymmetric branes in maximal supergravity theories define highest-weight orbits under the action of the symmetry group $G$.
These highest-weight orbits are single-charge orbits.
If not all the other long weights can be reached by an infinitesimal transformation, one can consider a two-charge state that is the sum of the initial state and the one that cannot be reached by the initial state. One can then compute the orbit of this 2-charge configuration. In case not all weights are reached one continues to consider 3-charge configurations etc. This procedure can be iterated until one has a configuration in which all the weights can be reached.
This strategy, that was used in \cite{Lu:1997bg} to compute the orbits for all standard branes, can be applied to non-standard branes as well.

All the different orbits of the bound states we are considering can be characterised in terms of invariants of $G$ \cite{Ferrara:1997ci}\,\footnote{We are not considering here more general  stationary configurations like the four-dimensional multi-black holes  of \cite{Bossard:2009my}, associated to  nilpotent orbits of the symmetry of the corresponding three-dimensional euclidean theory.}. For instance, the charge $T_M$ of a half-BPS string in six dimensions is a lightlike vector of $\text{SO}(5,5)$.  Therefore, the orbit for a single-charge configuration corresponds to the constraint $T^2=0$, while a two-charge configuration corresponds to $T^2 \neq 0$. As we have pointed out in section 3, the representations of the standard branes decompose under $H$ entirely into
the R-symmetry representations of the central charges, while for non-standard branes this decomposition gives these
 R-symmetry representations of the central charges plus additional representations (see Table \ref{nonstandardcentralcharge}). This implies that for standard branes the invariants of $G$ correspond to R-symmetry invariants of the central charge. These invariants characterise the amount of supersymmetry that the configuration preserves. For non-standard branes, instead, different invariants of $G$ may correspond to the same R-symmetry invariant when projected onto  the central charge. This means that multiple-charge configurations of non-standard branes, corresponding to different invariants and different orbits, can preserve the same amount of supersymmetry.

The aim of this section is to discuss  the orbits and invariants of non-standard branes. We will first review the case of standard branes in the first subsection, while the non-standard branes will be considered in the second subsection.

\begin{figure}
 \centering
\scalebox{0.4} 
{
\begin{pspicture}(0,-6.4306774)(8.284166,6.4306774)
\psline[linewidth=0.02cm](6.579583,-3.2348435)(2.3195832,-5.7548437)
\psline[linewidth=0.02cm](7.299583,-0.13484354)(5.819583,-2.8548436)
\psline[linewidth=0.02cm](5.179583,2.8051565)(6.559583,0.20515646)
\psline[linewidth=0.02cm](0.31958312,5.805156)(5.179583,3.2651565)
\usefont{T1}{ppl}{m}{n}
\rput(1.2305206,5.9551563){\huge \color{white}\psframebox[linewidth=0.02,fillstyle=solid,fillcolor=red]{1 0 0 0}}
\usefont{T1}{ppl}{m}{n}
\rput(1.3667706,-6.0448437){\huge \color{white}\psframebox[linewidth=0.02,fillstyle=solid,fillcolor=red]{0 0 0 -1}}
\usefont{T1}{ppl}{m}{n}
\rput(5.3667707,-3.0448434){\huge \color{white}\psframebox[linewidth=0.02,fillstyle=solid,fillcolor=red]{0 0 -1 1}}
\usefont{T1}{ppl}{m}{n}
\rput(6.775833,-0.044843547){\huge \color{white}\psframebox[linewidth=0.02,fillstyle=solid,fillcolor=red]{0 -1 1 0}}
\usefont{T1}{ppl}{m}{n}
\rput(5.381927,2.9551566){\huge \color{white}\psframebox[linewidth=0.02,fillstyle=solid,fillcolor=red]{-1 1 0 0}}
\usefont{T1}{ppl}{m}{n}
\rput{-27.708605}(-1.6921408,2.0807898){\rput(3.3323956,4.4901567){\Large $\alpha_{1}$}}
\usefont{T1}{ppl}{m}{n}
\rput{-63.806404}(2.1459281,6.349024){\rput(6.1323957,1.4701564){\Large $\alpha_{2}$}}
\usefont{T1}{ppl}{m}{n}
\rput{58.320164}(1.5029364,-6.0717173){\rput(6.1523957,-1.6698436){\Large $\alpha_{3}$}}
\usefont{T1}{ppl}{m}{n}
\rput{31.509905}(-1.9301298,-2.616664){\rput(3.6323957,-4.7098436){\Large $\alpha_{4}$}}
\end{pspicture}
}
\scalebox{0.4} 
{
\begin{pspicture}(0,-6.634271)(9.600729,6.634271)
\psline[linewidth=0.02cm](7.6477084,-3.2434373)(3.3877084,-5.7634373)
\psline[linewidth=0.02cm](8.367708,-0.1434373)(6.887708,-2.8634374)
\psline[linewidth=0.02cm](6.2477083,2.7965627)(7.6277084,0.1965627)
\psline[linewidth=0.02cm](1.3877083,5.7965627)(6.2477083,3.2565627)
\usefont{T1}{ppl}{m}{n}
\rput{31.509905}(-1.462984,-3.3696783){\rput(5.200521,-4.258437){\Large $T^{4}_{5}$}}
\usefont{T1}{ppl}{m}{n}
\rput(1.5222396,6.1265626){\huge \color{white}\psframebox[linewidth=0.04,fillstyle=solid,fillcolor=red]{$T_{1}$}}
\usefont{T1}{ppl}{m}{n}
\rput(6.3222394,3.1265626){\huge \color{white}\psframebox[linewidth=0.04,fillstyle=solid,fillcolor=red]{$T_{2}$}}
\usefont{T1}{ppl}{m}{n}
\rput(7.92224,-0.073437296){\huge \color{white}\psframebox[linewidth=0.04,fillstyle=solid,fillcolor=red]{$T_{3}$}}
\usefont{T1}{ppl}{m}{n}
\rput(3.1222396,-6.073437){\huge \color{white}\psframebox[linewidth=0.04,fillstyle=solid,fillcolor=red]{$T_{5}$}}
\usefont{T1}{ppl}{m}{n}
\rput(7.3222394,-3.0734372){\huge \color{white}\psframebox[linewidth=0.04,fillstyle=solid,fillcolor=red]{$T_{4}$}}
\usefont{T1}{ppl}{m}{n}
\rput{59.599396}(2.3118954,-6.9921584){\rput(7.220521,-1.4584373){\Large $T^{3}_{4}$}}
\usefont{T1}{ppl}{m}{n}
\rput{-61.73042}(2.0902603,7.262076){\rput(7.0805206,1.9015627){\Large $T^{2}_{3}$}}
\usefont{T1}{ppl}{m}{n}
\rput{-29.180836}(-1.9472893,2.4438004){\rput(3.6805208,4.9815626){\Large $T^{1}_{2}$}}
\end{pspicture}
}

\caption{\label{the5ofsl5} The Dynkin labels and the corresponding components of the ${\bf 5}$ of $\text{SL}(5,\mathbb{R})$. On the left-hand side of the figure, we write down the simple roots that are subtracted to each weight to get the weight below. On the right-hand side, we write down the corresponding generator in components.}
\end{figure}
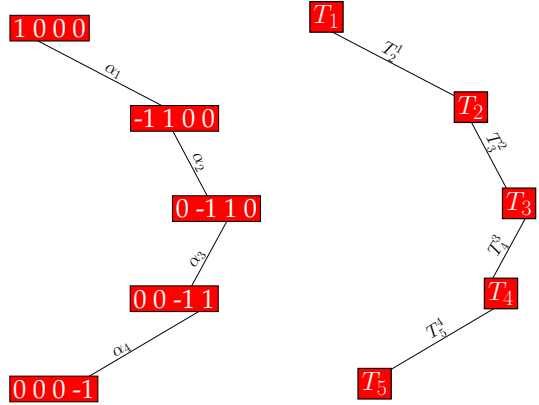

\subsection{Standard-brane orbits}

We will consider $D=7$ as a prototype example. The global symmetry group is $\text{SL}(5,\mathbb{R})$, and the strings correspond to  charges $T_{ M} $ in the ${\bf 5}$. Using the same component notation as done in section 2 for the
$\text{SL}(3,\mathbb{R})$ case, we associate charge components to weights as in Fig.~\ref{the5ofsl5}.
The representation clearly has only one dominant weight and all the weights have the same length. All the weights thus correspond to branes. The stabiliser of the highest weight orbit is generated by all the elements of the algebra that annihilate the highest weight. These are
the Cartan generators $H_{\alpha_{2}}$, $H_{\alpha_{3}}$ and $H_{\alpha_{4}}$, all the positive root vectors and all the negative root vectors that do not contain $\alpha_{1}$. The positive root vectors that do not contain $\alpha_1$, together with the negative root vectors and the Cartan stabilisers, generate the group $\text{SL}(4,\mathbb{R})$, while the remaining positive root vectors form the ${\bf 4}$ of this algebra.
 The orbit is therefore~\cite{Lu:1997bg}
  \begin{equation}
  \dfrac{\text{SL}(5,\mathbb{R})}{\text{SL}(4,\mathbb{R})\ltimes T_{\bf 4}} \label{theorbitofhw5ofsl5}
 \quad .
\end{equation}

\begin{figure}
 \centering
\scalebox{0.4} 
{
\begin{pspicture}(0,-9.430677)(13.672135,9.430677)
\definecolor{color35}{rgb}{0.996078431372549,0.996078431372549,0.996078431372549}
\definecolor{color34b}{rgb}{0.807843137254902,0.0,0.0}
\definecolor{color36b}{rgb}{0.8313725490196079,0.00784313725490196,0.00784313725490196}
\definecolor{color40b}{rgb}{0.8196078431372549,0.00392156862745098,0.00392156862745098}
\psline[linewidth=0.02cm](5.852396,-6.154843)(7.5723953,-8.814843)
\psline[linewidth=0.02cm](0.35239577,-3.254843)(5.1323953,-5.694843)
\psline[linewidth=0.02cm](8.352395,-3.194843)(6.692395,-5.7548428)
\psline[linewidth=0.02cm](1.9523957,-0.2148429)(6.892396,-2.694843)
\psline[linewidth=0.02cm](13.4123955,-0.2148429)(9.172396,-2.734843)
\psline[linewidth=0.02cm](3.5723956,-0.2148429)(1.8923957,-2.814843)
\psline[linewidth=0.02cm](0.9923958,2.805157)(2.8323956,0.18515709)
\psline[linewidth=0.02cm](9.132396,2.745157)(4.352396,0.2651571)
\psline[linewidth=0.02cm](6.772396,2.805157)(11.392395,0.2651571)
\psline[linewidth=0.02cm](7.412396,5.825157)(2.4923956,3.225157)
\psline[linewidth=0.02cm](5.772396,5.765157)(7.5723953,3.285157)
\psline[linewidth=0.02cm](8.452395,8.825157)(6.732396,6.305157)
\usefont{T1}{ppl}{m}{n}
\rput(8.054896,8.955156){\huge \color{color35}\psframebox[linewidth=0.02,fillstyle=solid,fillcolor=color34b]{0 0 1 0}}
\usefont{T1}{ppl}{m}{n}
\rput(1.3542707,-3.044843){\huge \color{color35}\psframebox[linewidth=0.02,fillstyle=solid,fillcolor=color36b]{1 0 -1 0}}
\usefont{T1}{ppl}{m}{n}
\rput(7.7848954,-9.044844){\huge \color{color35}\psframebox[linewidth=0.02,fillstyle=solid,fillcolor=color34b]{0 -1 0 0}}
\usefont{T1}{ppl}{m}{n}
\rput(7.7361455,2.955157){\huge \color{color35}\psframebox[linewidth=0.02,fillstyle=solid,fillcolor=color40b]{1 -1 0 1}}
\usefont{T1}{ppl}{m}{n}
\rput(3.0661457,-0.044842906){\huge \color{color35}\psframebox[linewidth=0.02,fillstyle=solid,fillcolor=color34b]{1 -1 1 -1}}
\usefont{T1}{ppl}{m}{n}
\rput(7.908958,-3.044843){\huge \color{color35}\psframebox[linewidth=0.02,fillstyle=solid,fillcolor=color34b]{-1 0 1 -1}}
\usefont{T1}{ppl}{m}{n}
\rput(6.166771,5.9551573){\huge \color{color35}\psframebox[linewidth=0.02,fillstyle=solid,fillcolor=color40b]{0 1 -1 1}}
\usefont{T1}{ppl}{m}{n}
\rput(1.3667709,2.955157){\huge \color{color35}\psframebox[linewidth=0.02,fillstyle=solid,fillcolor=color34b]{0 1 0 -1}}
\usefont{T1}{ppl}{m}{n}
\rput(12.178959,-0.044842906){\huge \color{color35}\psframebox[linewidth=0.02,fillstyle=solid,fillcolor=color34b]{-1 0 0 1}}
\usefont{T1}{ppl}{m}{n}
\rput(6.1270833,-6.0448427){\huge \color{color35}\psframebox[linewidth=0.02,fillstyle=solid,fillcolor=color34b]{-1 1 -1 0}}
\usefont{T1}{ptm}{m}{n}
\rput{-25.741508}(1.1330917,2.0318854){\rput(4.9728127,-1.444323){\Large $\alpha_{1}$}}
\usefont{T1}{ptm}{m}{n}
\rput{-54.618923}(9.14427,2.582438){\rput(7.0328126,-7.544323){\Large $\alpha_{2}$}}
\usefont{T1}{ptm}{m}{n}
\rput{-29.042452}(0.53184146,4.9662566){\rput(9.812813,1.4756771){\Large $\alpha_{1}$}}
\usefont{T1}{ptm}{m}{n}
\rput{-55.36516}(-0.2839217,2.491726){\rput(2.1928124,1.5356771){\Large $\alpha_{2}$}}
\usefont{T1}{ptm}{m}{n}
\rput{-25.25097}(2.1731179,0.9346147){\rput(3.1328125,-4.364323){\Large $\alpha_{1}$}}
\usefont{T1}{ptm}{m}{n}
\rput{-55.1361}(-0.39414895,7.757954){\rput(7.1928124,4.275677){\Large $\alpha_{2}$}}
\usefont{T1}{ptm}{m}{n}
\rput{56.056866}(-0.44232348,-7.9361887){\rput(7.1928124,-4.364323){\Large $\alpha_{3}$}}
\usefont{T1}{ptm}{m}{n}
\rput{30.824013}(0.79333687,-5.764919){\rput(10.812813,-1.424323){\Large $\alpha_{4}$}}
\usefont{T1}{ptm}{m}{n}
\rput{54.099735}(-0.57841253,-2.5143232){\rput(2.1328125,-1.804323){\Large $\alpha_{3}$}}
\usefont{T1}{ptm}{m}{n}
\rput{31.192688}(1.4824241,-2.7978384){\rput(5.7128124,1.2756771){\Large $\alpha_{4}$}}
\usefont{T1}{ptm}{m}{n}
\rput{57.646217}(9.693213,-2.5821402){\rput(7.1528125,7.535677){\Large $\alpha_{3}$}}
\usefont{T1}{ptm}{m}{n}
\rput{29.193901}(2.582943,-1.3653184){\rput(3.8728125,4.295677){\Large $\alpha_{4}$}}
\end{pspicture}
}
\scalebox{0.4} 
{
\begin{pspicture}(0,-9.724231)(14.200729,9.724468)
\definecolor{color2}{rgb}{0.996078431372549,0.996078431372549,0.996078431372549}
\definecolor{color1b}{rgb}{0.807843137254902,0.0,0.0}
\definecolor{color3b}{rgb}{0.8196078431372549,0.00392156862745098,0.00392156862745098}
\psline[linewidth=0.02cm](6.6877084,5.5923414)(1.8677083,3.3523417)
\psline[linewidth=0.02cm](6.1077085,5.8123417)(7.907708,3.3323417)
\psline[linewidth=0.02cm](1.3277084,2.8523417)(3.5477083,0.31234163)
\psline[linewidth=0.02cm](8.227708,2.6923416)(3.8877084,0.37234163)
\psline[linewidth=0.02cm](7.9877086,2.6323416)(12.287708,0.31234163)
\psline[linewidth=0.02cm](12.287708,-0.40765837)(8.327708,-2.6676583)
\psline[linewidth=0.02cm](3.5677083,-0.36765838)(8.187708,-2.6476583)
\psline[linewidth=0.02cm](3.9077084,-0.16765839)(1.5277083,-2.6476583)
\psline[linewidth=0.02cm](6.0477085,-6.327658)(7.6877084,-8.687658)
\psline[linewidth=0.02cm](8.187708,-3.2476585)(6.1677084,-5.827658)
\psline[linewidth=0.02cm](1.4477084,-3.3676584)(6.2677083,-5.787658)
\psline[linewidth=0.02cm](8.787708,8.872341)(7.0677085,6.3523417)
\usefont{T1}{ppl}{m}{n}
\rput(8.922239,9.002341){\huge \color{color2}\psframebox[linewidth=0.02,fillstyle=solid,fillcolor=color1b]{$T^{45}$}}
\usefont{T1}{ppl}{m}{n}
\rput(6.3222394,-5.9976583){\huge \color{color2}\psframebox[linewidth=0.02,fillstyle=solid,fillcolor=color3b]{$T^{13}$}}
\usefont{T1}{ppl}{m}{n}

\rput(6.5222397,6.0023417){\huge \color{color2}\psframebox[linewidth=0.02,fillstyle=solid,fillcolor=color3b]{$T^{35}$}}
\usefont{T1}{ppl}{m}{n}
\rput(8.32224,-2.9976585){\huge \color{color2}\psframebox[linewidth=0.02,fillstyle=solid,fillcolor=color1b]{$T^{14}$}}
\usefont{T1}{ppl}{m}{n}
\rput(1.7222396,-2.9976585){\huge \color{color2}\psframebox[linewidth=0.02,fillstyle=solid,fillcolor=color1b]{$T^{23}$}}
\usefont{T1}{ppl}{m}{n}
\rput(3.7222395,0.002341618){\huge \color{color2}\psframebox[linewidth=0.02,fillstyle=solid,fillcolor=color1b]{$T^{24}$}}
\usefont{T1}{ppl}{m}{n}
\rput(8.32224,3.0023415){\huge \color{color2}\psframebox[linewidth=0.02,fillstyle=solid,fillcolor=color1b]{$T^{25}$}}
\usefont{T1}{ppl}{m}{n}
\rput(1.9222395,3.0023415){\huge \color{color2}\psframebox[linewidth=0.02,fillstyle=solid,fillcolor=color3b]{$T^{34}$}}
\usefont{T1}{ppl}{m}{n}
\rput(7.92224,-8.997659){\huge \color{color2}\psframebox[linewidth=0.02,fillstyle=solid,fillcolor=color1b]{$T^{12}$}}
\usefont{T1}{ppl}{m}{n}
\rput(12.32224,0.002341618){\huge \color{color2}\psframebox[linewidth=0.02,fillstyle=solid,fillcolor=color1b]{$T^{15}$}}
\usefont{T1}{ppl}{m}{n}
\rput{27.923681}(1.5524259,-2.5679078){\rput(5.900521,1.8573416){\Large $T^{4}_{5}$}}
\usefont{T1}{ppl}{m}{n}
\rput{-52.832184}(0.017797528,2.9920733){\rput(2.9805207,1.4973416){\Large $T^{2}_{3}$}}
\usefont{T1}{ppl}{m}{n}
\rput{-53.010376}(-0.23808256,7.9188347){\rput(7.780521,4.2173414){\Large $T^{2}_{3}$}}
\usefont{T1}{ppl}{m}{n}
\rput{24.464548}(2.256619,-1.132641){\rput(3.7005208,4.6573415){\Large $T^{4}_{5}$}}
\usefont{T1}{ppl}{m}{n}
\rput{51.624146}(8.784463,-3.045522){\rput(7.5005207,7.5773416){\Large $T^{3}_{4}$}}
\usefont{T1}{ppl}{m}{n}
\rput{27.923681}(0.650509,-4.940226){\rput(10.220521,-1.1426584){\Large $T^{4}_{5}$}}
\usefont{T1}{ppl}{m}{n}
\rput{-27.623865}(1.3219821,2.7335584){\rput(6.180521,-1.3026584){\Large $T^{1}_{2}$}}
\usefont{T1}{ppl}{m}{n}
\rput{-28.579546}(0.6402201,5.389804){\rput(10.86052,1.4573417){\Large $T^{1}_{2}$}}
\usefont{T1}{ppl}{m}{n}
\rput{51.111923}(-0.65760094,-7.0285234){\rput(6.9805207,-4.182658){\Large $T^{3}_{4}$}}
\usefont{T1}{ppl}{m}{n}
\rput{-24.14912}(2.187732,1.4232564){\rput(4.380521,-4.3826585){\Large $T^{1}_{2}$}}
\usefont{T1}{ppl}{m}{n}
\rput{41.616947}(-0.37697127,-1.8918123){\rput(2.260521,-1.4226583){\Large $T^{3}_{4}$}}
\usefont{T1}{ppl}{m}{n}
\rput{-54.95811}(9.251431,2.8239539){\rput(7.300521,-7.4626584){\Large $T^{2}_{3}$}}
\end{pspicture}
}

\caption{\label{the10barofsl5} The Dynkin labels and the components of the ${\bf \overline{10}}$ of $\text{SL}(5,\mathbb{R})$. We denote the simple roots and the corresponding generators connecting the weights as explained in the caption of Fig. \ref{the5ofsl5}.}
\end{figure}
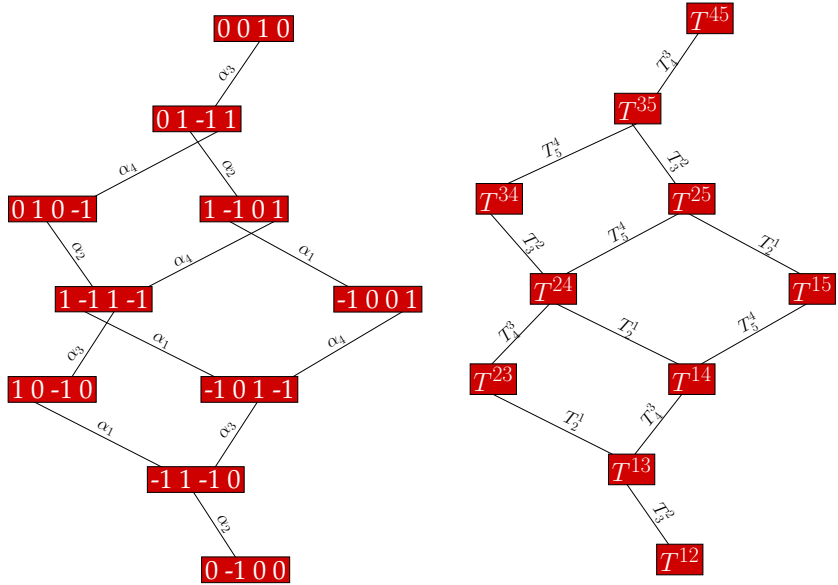

The charges $T^{MN}=-T^{NM}$ for the 0-branes are  in the ${\bf \overline{10}}$. The weights in terms of Dynkin labels and components are shown in Fig.~\ref{the10barofsl5}.
Again, there is only one dominant weight, and all the weights have the same length, and thus they all correspond to branes. The generators that annihilate the highest weight are  the Cartan's $H_{\alpha_{1}}$, $H_{\alpha_{2}}$ and $H_{\alpha_{4}}$, all the positive root vectors and the negative root vectors that do not contain $\alpha_3$. Altoghether, this generates the orbit~\cite{Lu:1997bg}
\begin{equation}
\dfrac{\text{SL}(5,\mathbb{R})}{\bigr(\text{SL}(3,\mathbb{R})\times \text{SL}(2,\mathbb{R})\bigl)\ltimes T^{({\bf 3,2})}}\,.
\end{equation}

\begin{figure}
\centering
\scalebox{0.6} 
{
\begin{pspicture}(0,-5.7914467)(17.505625,5.7914467)
\psbezier[linewidth=0.02,fillstyle=solid,fillcolor=blue,opacity=0.5](7.06,-5.7785535)(8.326644,-5.7814465)(10.780136,-0.9977477)(10.74,0.0014464753)(10.699863,1.0006407)(9.832819,2.4410713)(8.84,2.3414464)(7.847181,2.2418215)(8.1,1.5014465)(7.7,0.78144646)(7.3,0.061446477)(5.3121037,0.87740755)(5.32,-0.058553524)(5.3278966,-0.9945146)(5.7933564,-5.7756605)(7.06,-5.7785535)
\psbezier[linewidth=0.02,fillstyle=solid,fillcolor=yellow,opacity=0.5](7.14,5.7814465)(8.18,5.7814465)(10.7764635,0.756783)(10.68,-0.23855352)(10.583537,-1.23389)(9.44,-2.5985534)(8.94,-2.5385535)(8.44,-2.4785535)(8.24,-2.0585535)(7.78,-0.9985535)(7.32,0.061446477)(6.12105,-1.0707908)(5.54,-0.09855352)(4.9589505,0.8736837)(5.212335,2.547707)(5.62,3.6814466)(6.027665,4.815186)(6.1,5.7814465)(7.14,5.7814465)
\psdiamond[linewidth=0.02,dimen=outer](8.12,0.021446476)(2.0,3.6)
\psdots[dotsize=0.2](8.12,3.5814464)
\psdots[dotsize=0.2](6.12,0.021446476)
\psdots[dotsize=0.2](10.08,0.061446477)
\psdots[dotsize=0.2](8.1,-3.5185535)
\psdots[dotsize=0.2](9.1,-1.7785535)
\psdots[dotsize=0.2](7.1,-1.7785535)
\psdots[dotsize=0.2](7.14,1.8414465)
\psdots[dotsize=0.2](9.08,1.8414465)
\psline[linewidth=0.02cm](7.06,5.4014463)(8.14,3.5814464)
\psline[linewidth=0.02cm](8.1,-3.5385535)(7.16,-5.3785534)
\psdots[dotsize=0.2](7.06,5.4214463)
\psdots[dotsize=0.2](7.16,-5.3785534)
\usefont{T1}{ppl}{m}{n}
\rput(8.832812,3.5864465){\Large $\alpha_{3}$}
\usefont{T1}{ppl}{m}{n}
\rput(5.8428125,1.8464465){\Large $\alpha_{3}+\alpha_{4}$}
\usefont{T1}{ppl}{m}{n}
\rput(10.402813,1.8464465){\Large $\alpha_{2}+\alpha_{3}$}
\usefont{T1}{ppl}{m}{n}
\rput(4.0928126,0.046446476){\Large $\alpha_{2}+\alpha_{3}+\alpha_{4}$}
\usefont{T1}{ppl}{m}{n}
\rput(11.892813,0.066446476){\Large $\alpha_{1}+\alpha_{2}+\alpha_{3}$}
\usefont{T1}{ppl}{m}{n}
\rput(5.1628125,-1.7935535){\Large $\alpha_{2}+2\alpha_{3}+\alpha_{4}$}
\usefont{T1}{ppl}{m}{n}
\rput(11.702813,-1.7735535){\Large $\alpha_{1}+\alpha_{2}+\alpha_{3}+\alpha_{4}$}
\usefont{T1}{ppl}{m}{n}
\rput(10.812812,-3.5135536){\Large $\alpha_{1}+\alpha_{2}+2\alpha_{3}+\alpha_{4}$}
\usefont{T1}{ppl}{m}{n}
\rput(10.142813,-5.3535533){\Large $\alpha_{1}+2\alpha_{2}+2\alpha_{3}+\alpha_{4}$}
\pscircle[linewidth=0.02,dimen=outer,fillstyle=solid,fillcolor=red](7.08,5.4014463){0.2}
\pscircle[linewidth=0.02,dimen=outer,fillstyle=solid,fillcolor=red](7.16,-5.3785534){0.2}
\end{pspicture}

}
\caption{This figure describes the weights of the ${\bf \overline{10}}$ that can be reached by an infinitesimal $\text{SL}(5,\mathbb{R})$ transformation starting from the highest weight (yellow set) and from the lowest weight (blue set). The intersection of the two sets contains the weights associated to the conjunction stabilisers.
For each weight we write  its distance  from the highest weight in terms of simple roots. 
 \label{10barhighestlowest}}
\end{figure}
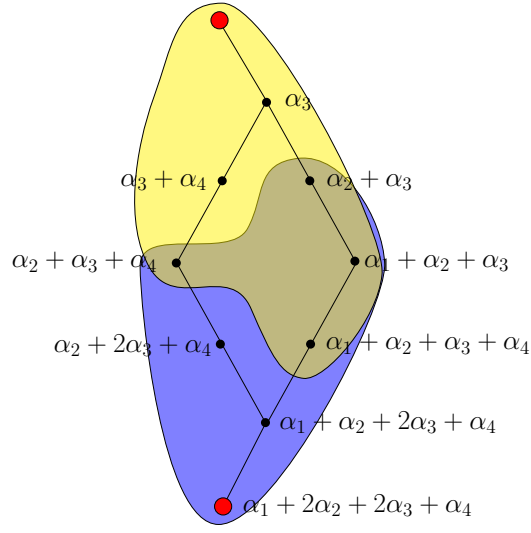

By looking at Figs. \ref{the5ofsl5} and \ref{the10barofsl5}, one notices that while in the case of the ${\bf  5}$  any weight can be reached starting by any other weight by the action of a given generator, in the case of the ${\bf \overline{10}}$ this is no longer true: if one considers any weight in the representation, there are always three weights that are not connected to it by transformations in the algebra. In particular, if one considers the highest weight, one can see that the weights $\boxed{1 \ 0 \ -1 \ 0}$, $\boxed{-1 \ 1 \ -1 \ 0}$ and $\boxed{0 \ -1 \ 0 \ 0}$  in fig.~\ref{the10barofsl5} are not connected by transformations in the algebra.
This can be easily seen by noticing that the difference between the weight $\boxed{1 \ 0 \ -1 \ 0}$ and the highest weight is $\alpha_2 +2 \alpha_3 + \alpha_4$ which is not a root.
 This is also easy to understand in terms of components: the highest weight corresponds to the charge $T^{45}$, and the three charges $T^{12}$, $T^{13}$ and $T^{23}$ are not connected because an infinitesimal transformation cannot change both indices. One can then compute the orbit of a 2-charge configuration, which for instance we choose to be $T^{45}+ T^{12}$. In the 2-charge case, the generators that stabilise the orbit are not only the common stabilisers of both weights, but also those generators that take the two weights to the same weight with opposite sign, so that the overall transformation vanishes. This can be seen in Fig. \ref{10barhighestlowest}, which shows that  the components $T^{14}$, $T^{15}$, $T^{24}$ and $T^{25}$ are connected  to both the highest weight and the lowest weight by infinitesimal transformations. In general,  we call such generators the ``conjunction'' stabilisers.

The 2-charge orbit is determined as follows~\cite{Lu:1997bg}.
 The common stabilisers are the Cartan generators  $H_{\alpha_{1}}$ and $H_{\alpha_{4}}$ and the root vectors
\begin{equation}
  E_{\pm\alpha_{1}}\ , \ \ E_{\pm\alpha_{4}}\ ,\ \ E_{-\alpha_{2}}\ ,\ \ E_{\alpha_{3}}\ ,\ \  E_{\alpha_{3}+\alpha_{4}}\ ,\ \ E_{-\alpha_{1}-\alpha_{2}} \ ,
\end{equation}
 while the conjunction stabilisers are
\begin{eqnarray}
& & E_{\alpha_{1}+\alpha_{2}+\alpha_{3}}-E_{-\alpha_{2}-\alpha_{3}-\alpha_{4}} \ , \ \  E_{\alpha_{1}+\alpha_{2}+\alpha_{3}+\alpha_{4}}-E_{-\alpha_{2}-\alpha_{3}}\ ,\nonumber \\ & & E_{\alpha_{2}+\alpha_{3}+\alpha_{4}}-E_{-\alpha_{1}-\alpha_{2}-\alpha_{3}}\ , \ \  E_{\alpha_{2}+\alpha_{3}}-E_{-\alpha_{1}-\alpha_{2}-\alpha_{3}-\alpha_{4}} \ \ .
\end{eqnarray}
To extract the semisimple part  we make the identifications
\begin{eqnarray}
&& E_{\alpha_{1}+\alpha_{2}+\alpha_{3}}-E_{-\alpha_{2}-\alpha_{3}-\alpha_{4}}\rightarrow E_{\frac{\alpha_{1}-\alpha_{4}}{2}}\ ,\nonumber \\
&&E_{\alpha_{1}+\alpha_{2}+\alpha_{3}+\alpha_{4}}-E_{-\alpha_{2}-\alpha_{3}}\rightarrow E_{\frac{\alpha_{1}+\alpha_{4}}{2}}\ ,\nonumber \\
&& E_{\alpha_{2}+\alpha_{3}+\alpha_{4}}-E_{-\alpha_{1}-\alpha_{2}-\alpha_{3}}\rightarrow E_{\frac{-\alpha_{1}+\alpha_{4}}{2}}\ ,\nonumber \\
&& E_{\alpha_{2}+\alpha_{3}}-E_{-\alpha_{1}-\alpha_{2}-\alpha_{3}-\alpha_{4}}\rightarrow E_{-\frac{\alpha_{1}+\alpha_{4}}{2}} \ .
\end{eqnarray}
Defining $\beta_{1}=\frac{\alpha_{1}-\alpha_{4}}{2}$ and $\beta_{2}=\alpha_{4}$ one recognises that the conjunction stabilisers together with the stabilizing roots $\pm\alpha_{1}$ and $\pm\alpha_{4}$ and the Cartan generators $
H_{\beta_{1}}= H_{\frac{\alpha_{1}-\alpha_{4}}{2}}=H_{\alpha_{1}}-H_{\alpha_{4}}$, $H_{\beta_{2}}=H_{\alpha_{4}}$
generate an algebra $SO(2,3)$ with simple roots $\beta_{1}$ and $\beta_{2}$; the rest of the stabilising roots reorganise
themselves in the representation ${\bf 4}$ of this group with highest weight $-\alpha_{2}$. Thus, the two-charge orbit for the one forms in $D=7$ (coupling to 0-branes)  is the 10-dimensional coset
\begin{equation}
 \dfrac{Sl(5,\mathbb{R})}{SO(2,3)\ltimes T^{\bf 4}}.
\end{equation}
Given that all the weights can be reached in any 2-charge orbit, there are no configurations with 3 charges in this case.

The fact that there is only a one-charge orbit in the ${\bf 5}$, while there is also a 2-charge orbit in the ${\bf \overline{10}}$ can be understood in terms of invariants: for $T_M$ in the  ${\bf 5}$, there is no non-trivial contraction with the invariant tensor $\epsilon^{M_1 ...M_5}$ that one can write, while for $T^{MN}$ one can construct
  \begin{equation}\label{inv}
T^{MN} T^{PQ} \epsilon_{MNPQR} \quad .
\end{equation}
For the highest weight orbit, for which only one component of the charge is turned on, this quantity is clearly vanishing, while it is not vanishing for the 2-charge orbit.

We now discuss the amount of supersymmetry that these orbits preserve. As we discussed  in section 3, the representations of $\text{SL}(5,\mathbb{R})$ for standard branes decompose under the R-symmetry $\text{SO}(5)$ as in the first line of eq. \eqref{standardTQnonstandardTRQ}, to give entirely the
central charges $Q$.  This means that the $\text{SL}(5,\mathbb{R})$ invariants in terms of $T$ are identical to the R-symmetry invariants in terms of the central charges $Q$. In the case of the ${\bf 5}$ of  $\text{SL}(5,\mathbb{R})$, there is only one orbit, and therefore any charge in the ${\bf 5}$ corresponds to a half-BPS configuration. For the ${\bf \overline{10}}$, the invariant of $\text{SL}(5,\mathbb{R})$, given in eq.~\eqref{inv}, is mapped to the same invariant of $\text{SO}(5)$ written in terms of the central charges $Q_{MN}$, and thus the single-charge and the 2-charge orbits 
 preserve a different amount of supersymmetry, namely  1/2 and 1/4 respectively.

This analysis is completely general. For instance, in $D=4$ the 0-brane orbits are classified in terms of a quartic invariant~\cite{Ferrara:1997ci}
\begin{equation}
I_4 = d^{MNPQ} T_M T_N T_P T_Q  \quad ,
\end{equation}
where the index $M$ denotes the ${\bf 56}$ of $\text{E}_{7(7)}$ and $d^{MNPQ}$ is the invariant tensor of $E_{7(7)}$ in the fully symmetric product of 4 ${\bf 56}$'s. The highest weight orbit is given by
\begin{equation}
 I_4 = \frac{\partial I_4}{\partial T_M} = \frac{\partial^2 I_4}{\partial T_M \partial T_N} =0 \quad ,
\end{equation}
and it preserves 16 supercharges. The 2-charge orbit is given by the constraints
\begin{equation}
 I_4 =  \frac{\partial I_4}{\partial T_M} = 0 \ \ \frac{\partial^2 I_4}{\partial T_M \partial T_N} \neq0 \quad ,
\end{equation}
and it preserves 8 supercharges. The 3-charge orbit is given by
\begin{equation}
 I_4 = 0\ \  \frac{\partial I_4}{\partial T_M} \neq 0 \quad ,
\end{equation}
and it preserves 4 supercharges. Finally, the case $I_4 \neq 0$ can either give a 4-charge orbit, preserving again 4 supercharges, or a dyonic orbit, preserving no supersymmetry at all.

A complete summary of orbits and invariants in maximal supergravity theories can be found in \cite{Borsten:2010aa}.
This finishes our discussion of the standard brane orbits. In the next subsection we will consider the non-standard-brane orbits, and the supersymmetry that they preserve.

\subsection{Non-standard-brane orbits}

The single-charge orbits for non-standard branes have been derived in \cite{Bergshoeff:2012ex}. Here we consider the multiple-charge orbits for non-standard branes. Again, we focus on the seven-dimensional case and we consider in particular the tensor 5-branes, with charges $T^{MN}=T^{NM}$ in the ${\bf \overline{15}}$.
The Dynkin labels of the weights, and the corresponding components, are shown in Fig.~\ref{the15barofsl5}. The five brane-charges $T^{MM}$ correspond to the long weights, which are as usual painted in red in the figure. The other ten weights are exactly the weights of  the ${\bf \overline{10}}$, as one can see by looking at Fig.~\ref{the10barofsl5}. The highest weight of the ${\bf \overline{10}}$ is $\boxed{0\ 0\ 1 \ 0}$, and indeed this weight is present as a dominant weight in Fig.~\ref{the15barofsl5}.

\begin{figure}
 \centering
\scalebox{0.3} 
{
\begin{pspicture}(0,-16.430677)(15.284167,16.430677)
\definecolor{color189}{rgb}{0.996078431372549,0.996078431372549,0.996078431372549}
\definecolor{color191b}{rgb}{0.807843137254902,0.0,0.0}
\psline[linewidth=0.02cm](8.379583,7.8051558)(4.1595836,4.3251557)
\usefont{T1}{ppl}{m}{n}
\rput{39.154793}(5.2771573,-2.4566276){\rput(6.052396,6.210156){\Large $\alpha_{4}$}}
\usefont{T1}{ppl}{m}{n}
\rput{35.41034}(1.0519768,-7.1933784){\rput(11.752396,-1.9298441){\Large $\alpha_{4}$}}
\psline[linewidth=0.02cm](14.559584,-0.27484417)(9.959583,-3.5148442)
\psline[linewidth=0.02cm](8.099584,-12.174844)(12.939584,-15.754844)
\usefont{T1}{ppl}{m}{n}
\rput{-37.808403}(11.027237,3.8620942){\rput(11.112396,-14.149844){\Large $\alpha_{1}$}}
\psline[linewidth=0.02cm](3.4795833,-8.134844)(8.159583,-11.734844)
\psline[linewidth=0.02cm](7.4395833,-8.234844)(8.999583,-11.854844)
\usefont{T1}{ppl}{m}{n}
\rput{-39.194077}(7.7066684,1.7082552){\rput(6.2123957,-9.949844){\Large $\alpha_{1}$}}
\usefont{T1}{ppl}{m}{n}
\rput{-65.67299}(14.301894,1.7830572){\rput(8.492395,-10.169845){\Large $\alpha_{2}$}}
\psline[linewidth=0.02cm](2.5595834,-4.0748444)(4.2195835,-7.774844)
\psline[linewidth=0.02cm](1.8395833,-4.0548444)(9.119583,-7.6948442)
\psline[linewidth=0.02cm](9.099584,-3.9948442)(8.259583,-7.774844)
\usefont{T1}{ppl}{m}{n}
\rput{-24.752228}(2.6807318,1.5588244){\rput(4.852396,-5.309844){\Large $\alpha_{1}$}}
\usefont{T1}{ppl}{m}{n}
\rput{-63.836166}(7.811928,-0.11656395){\rput(3.7723958,-6.309844){\Large $\alpha_{2}$}}
\usefont{T1}{ppl}{m}{n}
\rput{76.59314}(0.89201146,-12.707747){\rput(8.452395,-5.769844){\Large $\alpha_{3}$}}
\psline[linewidth=0.02cm](0.9195833,-0.27484417)(2.2795832,-3.654844)
\psline[linewidth=0.02cm](3.2595832,-0.25484416)(7.7195835,-3.5348442)
\psline[linewidth=0.02cm](4.8795834,-0.23484416)(3.4395833,-3.594844)
\usefont{T1}{ppl}{m}{n}
\rput{62.927948}(0.024187824,-4.577653){\rput(3.712396,-2.249844){\Large $\alpha_{3}$}}
\usefont{T1}{ppl}{m}{n}
\rput{-37.051865}(2.3561027,3.1729453){\rput(5.872396,-1.9098442){\Large $\alpha_{1}$}}
\usefont{T1}{ppl}{m}{n}
\rput{-69.30491}(3.0714183,0.46543965){\rput(1.8323958,-1.9698441){\Large $\alpha_{2}$}}
\psline[linewidth=0.02cm](3.2395833,3.8051558)(1.8995833,0.18515584)
\psline[linewidth=0.02cm](2.4995832,3.8251557)(4.099583,0.20515585)
\psline[linewidth=0.02cm](7.6595836,3.7651558)(12.599584,0.20515585)
\psline[linewidth=0.02cm](9.959583,3.7451558)(5.7395835,0.26515585)
\usefont{T1}{ppl}{m}{n}
\rput{67.623085}(3.1497402,-0.9611019){\rput(2.2523959,1.8901558){\Large $\alpha_{3}$}}
\usefont{T1}{ppl}{m}{n}
\rput{39.154793}(3.0684018,-4.365969){\rput(7.6323957,2.1501558){\Large $\alpha_{4}$}}
\usefont{T1}{ppl}{m}{n}
\rput{-64.33946}(0.2782237,4.344204){\rput(3.5523958,1.9701558){\Large $\alpha_{2}$}}
\usefont{T1}{ppl}{m}{n}
\rput{-35.568764}(1.086935,6.6904535){\rput(10.932396,1.6701559){\Large $\alpha_{1}$}}
\psline[linewidth=0.02cm](4.6595836,7.8251557)(3.2595832,4.285156)
\psline[linewidth=0.02cm](6.9195833,7.8051558)(8.539583,4.205156)
\usefont{T1}{ppl}{m}{n}
\rput{66.949005}(7.606078,0.19922464){\rput(3.6123958,5.870156){\Large $\alpha_{3}$}}
\usefont{T1}{ppl}{m}{n}
\rput{-67.453}(-0.37120062,11.02584){\rput(8.032395,5.810156){\Large $\alpha_{2}$}}
\psline[linewidth=0.02cm](9.7195835,11.725156)(5.519583,8.305156)
\psline[linewidth=0.02cm](9.039583,11.825156)(7.8395834,8.185156)
\usefont{T1}{ppl}{m}{n}
\rput{41.565414}(8.206853,-2.1624763){\rput(6.912396,9.750155){\Large $\alpha_{4}$}}
\usefont{T1}{ppl}{m}{n}
\rput{65.48303}(13.0030155,-1.8401575){\rput(7.892396,9.2101555){\Large $\alpha_{3}$}}
\psline[linewidth=0.02cm](13.679584,15.805156)(9.639584,12.285156)
\usefont{T1}{ppl}{m}{n}
\rput{41.356895}(11.868701,-3.863497){\rput(11.012396,13.810156){\Large $\alpha_{4}$}}
\usefont{T1}{ppl}{m}{n}
\rput(12.634114,15.955155){\huge \color{color189}\psframebox[linewidth=0.02,fillstyle=solid,fillcolor=color191b]{0 0 0 2}}
\usefont{T1}{ppl}{m}{n}
\rput(9.175834,-12.044845){\huge \psframebox[linewidth=0.02,fillstyle=solid]{0 -1 0 0}}
\usefont{T1}{ppl}{m}{n}
\rput(7.711927,-8.044845){\huge \psframebox[linewidth=0.02,fillstyle=solid]{-1 1 -1 0}}
\usefont{T1}{ppl}{m}{n}
\rput(4.5733333,-8.044845){\huge \color{color189}\psframebox[linewidth=0.02,fillstyle=solid,fillcolor=color191b]{2 -2 0 0}}
\usefont{T1}{ppl}{m}{n}
\rput(8.702865,-3.844844){\huge \psframebox[linewidth=0.02,fillstyle=solid]{-1 0 1 -1}}
\usefont{T1}{ppl}{m}{n}
\rput(1.3758334,-0.044844158){\huge \color{color189}\psframebox[linewidth=0.02,fillstyle=solid,fillcolor=color191b]{0 2 -2 0}}
\usefont{T1}{ppl}{m}{n}
\rput(4.481458,-0.044844158){\huge \psframebox[linewidth=0.02,fillstyle=solid]{1 -1 1 -1}}
\usefont{T1}{ppl}{m}{n}
\rput(8.751458,3.9551558){\huge \psframebox[linewidth=0.02,fillstyle=solid]{1 -1 0 1}}
\usefont{T1}{ppl}{m}{n}
\rput(3.0267708,3.9551558){\huge \psframebox[linewidth=0.02,fillstyle=solid]{0 1 0 -1}}
\usefont{T1}{ppl}{m}{n}
\rput(2.760521,-3.844844){\huge \psframebox[linewidth=0.02,fillstyle=solid]{1 0 -1 0}}
\usefont{T1}{ppl}{m}{n}
\rput(4.3641148,7.955156){\huge \color{color189}\psframebox[linewidth=0.02,fillstyle=solid,fillcolor=color191b]{0 0 2 -2}}
\usefont{T1}{ppl}{m}{n}
\rput(7.3667707,7.955156){\huge \psframebox[linewidth=0.02,fillstyle=solid]{0 1 -1 1}}
\usefont{T1}{ppl}{m}{n}
\rput(8.645833,11.955155){\huge \psframebox[linewidth=0.02,fillstyle=solid]{0 0 1 0}}
\usefont{T1}{ppl}{m}{n}
\rput(13.372865,-0.044844158){\huge \psframebox[linewidth=0.02,fillstyle=solid]{-1 0 0 1}}
\usefont{T1}{ppl}{m}{n}
\rput(13.781927,-16.044844){\huge \color{color189}\psframebox[linewidth=0.02,fillstyle=solid,fillcolor=color191b]{-2 0 0 0}}
\end{pspicture}
}
\scalebox{0.3} 
{
\begin{pspicture}(0,-16.488802)(16.22073,16.488802)
\definecolor{color283}{rgb}{0.996078431372549,0.996078431372549,0.996078431372549}
\definecolor{color285b}{rgb}{0.807843137254902,0.0,0.0}
\definecolor{color27}{rgb}{0.00392156862745098,0.00392156862745098,0.00392156862745098}
\definecolor{color41b}{rgb}{0.7254901960784313,0.0,0.0}
\psline[linewidth=0.02cm](7.6077085,7.751094)(3.3877084,4.271094)
\usefont{T1}{ppl}{m}{n}
\rput{39.154793}(5.0696917,-1.9813921){\rput(5.280521,6.1560936){\Large $T_5^4$}}
\psline[linewidth=0.02cm](2.8277082,-4.148906)(8.387709,-7.668906)
\usefont{T1}{ppl}{m}{n}
\rput{35.41034}(0.9237371,-6.7797995){\rput(11.040521,-1.9239062){\Large $T_5^4$}}
\psline[linewidth=0.02cm](13.847709,-0.26890624)(9.247708,-3.5089064)
\psline[linewidth=0.02cm](9.467709,-12.188907)(14.307709,-15.768907)
\usefont{T1}{ppl}{m}{n}
\rput{-37.808403}(11.323074,4.697834){\rput(12.480521,-14.163906){\Large $T_2^1$}}
\psline[linewidth=0.02cm](4.927708,-8.128906)(9.607708,-11.728907)
\psline[linewidth=0.02cm](7.947708,-8.188907)(9.507709,-11.808907)
\usefont{T1}{ppl}{m}{n}
\rput{-39.194077}(8.02873,2.6247327){\rput(7.660521,-9.943906){\Large $T_2^1$}}
\usefont{T1}{ppl}{m}{n}
\rput{-65.67299}(14.558841,2.27308){\rput(9.000521,-10.123906){\Large $T_3^2$}}
\psline[linewidth=0.02cm](2.8349233,-4.031556)(4.7579303,-7.6019406)
\psline[linewidth=0.02cm](9.087708,-3.9489062)(8.247708,-7.728906)
\usefont{T1}{ppl}{m}{n}
\rput{-30.78627}(3.6775713,2.0713391){\rput(5.5605206,-5.623906){\Large $T_2^1$}}
\usefont{T1}{ppl}{m}{n}
\rput{-59.692616}(7.447431,0.60089105){\rput(4.207341,-6.170141){\Large $T_3^2$}}
\usefont{T1}{ppl}{m}{n}
\rput{76.59314}(0.9275766,-12.660911){\rput(8.440521,-5.723906){\Large $T_4^3$}}
\psline[linewidth=0.02cm](1.4277084,-0.18890625)(2.7877083,-3.5689063)
\psline[linewidth=0.02cm](4.427708,-0.20890625)(8.887709,-3.4889061)
\psline[linewidth=0.02cm](4.847708,-0.12890625)(3.4077084,-3.4889061)
\usefont{T1}{ppl}{m}{n}
\rput{62.927948}(0.10115026,-4.4915457){\rput(3.6805208,-2.1439064){\Large $T_4^3$}}
\usefont{T1}{ppl}{m}{n}
\rput{-37.051865}(2.5642788,3.8860598){\rput(7.0405207,-1.8639063){\Large $T_2^1$}}
\usefont{T1}{ppl}{m}{n}
\rput{-69.30491}(3.319582,0.99634534){\rput(2.3405209,-1.8839062){\Large $T_3^2$}}
\psline[linewidth=0.02cm](3.2277083,3.8510938)(1.8877083,0.23109375)
\psline[linewidth=0.02cm](3.0277083,3.8910937)(4.6277084,0.27109376)
\psline[linewidth=0.02cm](8.867708,3.8110938)(13.807709,0.25109375)
\psline[linewidth=0.02cm](9.387709,3.7510939)(5.1677084,0.27109376)
\usefont{T1}{ppl}{m}{n}
\rput{67.623085}(3.1848648,-0.92167175){\rput(2.2405207,1.9360938){\Large $T_4^3$}}
\usefont{T1}{ppl}{m}{n}
\rput{39.154793}(2.9437325,-4.003544){\rput(7.0605206,2.1560938){\Large $T_5^4$}}
\usefont{T1}{ppl}{m}{n}
\rput{-64.33946}(0.5182157,4.8576274){\rput(4.0805206,2.0360937){\Large $T_3^2$}}
\usefont{T1}{ppl}{m}{n}
\rput{-35.568764}(1.2856282,7.401767){\rput(12.140521,1.7160938){\Large $T_2^1$}}
\psline[linewidth=0.02cm](4.6477084,7.8710938)(3.2477083,4.331094)
\psline[linewidth=0.02cm](7.5077085,7.9110937)(9.127708,4.311094)
\usefont{T1}{ppl}{m}{n}
\rput{-67.453}(-0.10642798,11.634328){\rput(8.620521,5.916094){\Large $T_3^2$}}
\usefont{T1}{ppl}{m}{n}
\rput{66.949005}(7.6411233,0.23810239){\rput(3.6005208,5.916094){\Large $T_4^3$}}
\psline[linewidth=0.02cm](9.187708,11.811093)(4.9877086,8.391094)
\psline[linewidth=0.02cm](9.027708,11.871094)(7.8277082,8.231093)
\usefont{T1}{ppl}{m}{n}
\rput{41.565414}(8.129944,-1.7879512){\rput(6.380521,9.836094){\Large $T_5^4$}}
\usefont{T1}{ppl}{m}{n}
\rput{65.48303}(13.037864,-1.8024778){\rput(7.880521,9.256094){\Large $T_4^3$}}
\psline[linewidth=0.02cm](13.667708,15.851093)(9.627708,12.331094)
\usefont{T1}{ppl}{m}{n}
\rput{41.356895}(11.896092,-3.8441942){\rput(11.000521,13.856093){\Large $T_5^4$}}
\usefont{T1}{ppl}{m}{n}
\rput(13.23224,16.001093){\huge \color{color283}\psframebox[linewidth=0.02,fillstyle=solid,fillcolor=color285b]{$T^{55}$}}
\usefont{T1}{ppl}{m}{n}
\rput(9.632239,-11.998906){\huge \psframebox[linewidth=0.02,fillstyle=solid]{$T^{12}$}}
\usefont{T1}{ppl}{m}{n}
\rput(8.19224,-7.998906){\huge \psframebox[linewidth=0.02,fillstyle=solid]{$T^{13}$}}
\usefont{T1}{ppl}{m}{n}
\rput(5.0322394,-7.998906){\huge \color{color283}\psframebox[linewidth=0.02,linecolor=color27,fillstyle=solid,fillcolor=color285b]{$T^{22}$}}
\usefont{T1}{ppl}{m}{n}
\rput(9.03224,-3.7989063){\huge \psframebox[linewidth=0.02,fillstyle=solid]{$T^{14}$}}
\usefont{T1}{ppl}{m}{n}
\rput(1.8322396,0.00109375){\huge \color{color283}\psframebox[linewidth=0.02,fillstyle=solid,fillcolor=color285b]{$T^{33}$}}
\usefont{T1}{ppl}{m}{n}
\rput(4.8322396,0.00109375){\huge \psframebox[linewidth=0.02,fillstyle=solid]{$T^{24}$}}
\usefont{T1}{ppl}{m}{n}
\rput(9.23224,4.001094){\huge \psframebox[linewidth=0.02,fillstyle=solid]{$T^{25}$}}
\usefont{T1}{ppl}{m}{n}
\rput(3.4322395,4.001094){\huge \psframebox[linewidth=0.02,fillstyle=solid]{$T^{34}$}}
\usefont{T1}{ppl}{m}{n}
\rput(3.2322395,-3.7989063){\huge \psframebox[linewidth=0.02,fillstyle=solid]{$T^{23}$}}
\usefont{T1}{ppl}{m}{n}
\rput(4.8322396,8.001094){\huge \color{color283}\psframebox[linewidth=0.02,fillstyle=solid,fillcolor=color41b]{$T^{44}$}}
\usefont{T1}{ppl}{m}{n}
\rput(7.8322396,8.001094){\huge \psframebox[linewidth=0.02,fillstyle=solid]{$T^{35}$}}
\usefont{T1}{ppl}{m}{n}
\rput(9.23224,12.001094){\huge \psframebox[linewidth=0.02,fillstyle=solid]{$T^{45}$}}
\usefont{T1}{ppl}{m}{n}
\rput(13.832239,0.00109375){\huge \psframebox[linewidth=0.02,fillstyle=solid]{$T^{15}$}}
\usefont{T1}{ppl}{m}{n}
\rput(14.23224,-15.998906){\huge \color{color283}\psframebox[linewidth=0.02,fillstyle=solid,fillcolor=color285b]{$T^{11}$}}
\end{pspicture}
}

\caption{\label{the15barofsl5} The Dynkin labels and the components of the ${\bf \overline{15}}$ of $\text{SL}(5,\mathbb{R })$.  We denote the simple roots and the corresponding generators connecting the weights as explained in the caption of Fig. \ref{the5ofsl5}.}
\end{figure}

One can compute the orbits exactly as for the standard branes. The highest-weight orbit is the same as the highest-weight orbit of the ${\bf 5}$, which is given in eq.~\eqref{theorbitofhw5ofsl5}. The fact that these two highest-weight orbits coincide is not surprising since the weights that can be reached from the highest weight of the ${\bf \overline{15}}$ form the ${\bf \overline{5}}$ of $\text{SL}(5,\mathbb{R})$. In components, this means that infinitesimally one can only transform one of the two indices.

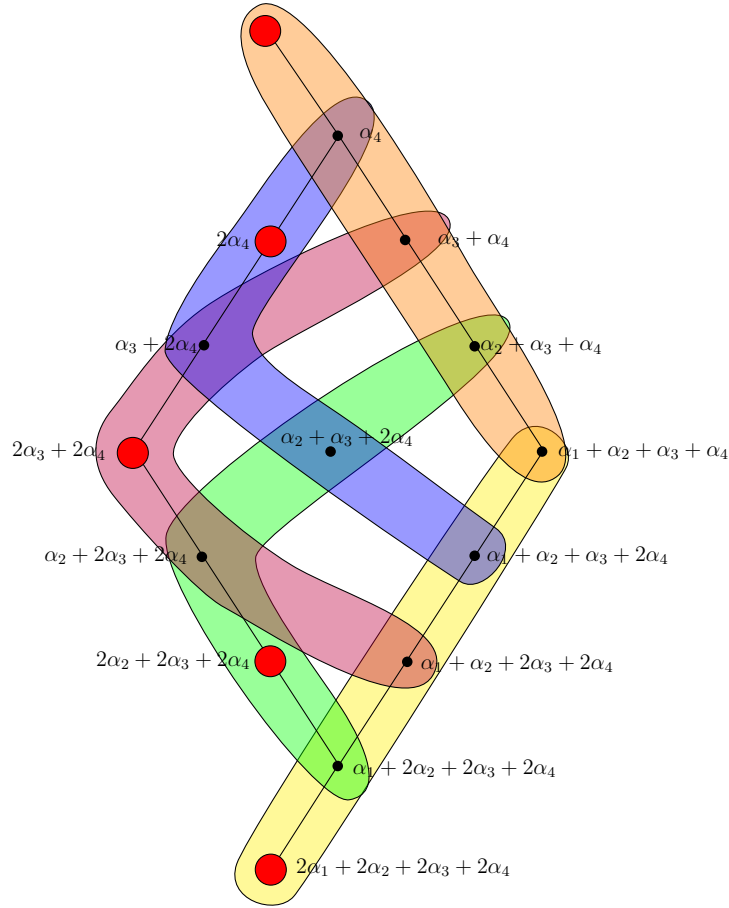
\begin{figure}
 \centering

\scalebox{0.7} 
{
\begin{pspicture}(0,-10.076092)(19.423656,10.076092)
\psbezier[linewidth=0.02,fillstyle=solid,fillcolor=yellow,opacity=0.4](13.744595,0.8460917)(13.524594,0.38609168)(8.904594,-6.9339085)(8.564594,-7.313908)(8.224594,-7.693908)(7.2500324,-7.3801284)(7.4845943,-6.693908)(7.719156,-6.007688)(12.291435,0.95848805)(12.764594,1.4460917)(13.237753,1.9336953)(13.964594,1.3060917)(13.744595,0.8460917)
\psbezier[linewidth=0.02,fillstyle=solid,fillcolor=green,opacity=0.4](9.624594,-5.4739084)(8.951031,-5.5923204)(5.9921303,-1.4422174)(6.144594,-0.45390832)(6.297058,0.5344007)(12.008481,4.393749)(12.624594,3.6460917)(13.240707,2.8984342)(7.884594,-0.27390832)(7.844594,-0.87390834)(7.804594,-1.4739083)(9.024594,-3.0339084)(9.484594,-3.7539084)(9.944594,-4.4739084)(10.298158,-5.3554964)(9.624594,-5.4739084)
\psbezier[linewidth=0.02,fillstyle=solid,fillcolor=purple,opacity=0.4](5.104594,1.9260917)(5.5045943,2.4260917)(6.046441,3.3110914)(6.904594,3.9060917)(7.7627473,4.501092)(11.104594,6.226092)(11.504594,5.5660915)(11.904594,4.9060917)(8.664594,3.4860916)(7.9245944,2.9460917)(7.184594,2.4060917)(6.220324,1.485968)(6.284594,1.0660917)(6.348864,0.64621544)(7.6049094,-0.45083404)(8.524594,-1.0139083)(9.444279,-1.5769826)(11.744595,-2.2139084)(11.204595,-3.1139083)(10.664594,-4.0139084)(7.744594,-2.0739083)(7.244594,-1.8139083)(6.744594,-1.5539083)(5.5845942,-0.2539083)(5.144594,0.3060917)(4.704594,0.86609167)(4.704594,1.4260917)(5.104594,1.9260917)
\psbezier[linewidth=0.02,fillstyle=solid,fillcolor=blue,opacity=0.4](10.004594,7.766092)(10.644439,7.092736)(7.704594,3.8660917)(7.784594,3.3660917)(7.864594,2.8660917)(8.595051,2.4245338)(9.444594,1.8460916)(10.294138,1.2676495)(11.864594,0.06609169)(12.404594,-0.3139083)(12.944594,-0.69390833)(12.184594,-1.5939083)(11.744595,-1.3339083)(11.304594,-1.0739083)(6.0930114,2.4480517)(6.124594,3.0660918)(6.156177,3.6841316)(6.9845943,4.826092)(7.4245944,5.4060917)(7.864594,5.9860916)(9.364749,8.439447)(10.004594,7.766092)
\psbezier[linewidth=0.02,fillstyle=solid,fillcolor=orange,opacity=0.4](7.904594,9.626092)(8.804594,10.066092)(11.304594,5.726092)(11.864594,4.746092)(12.424594,3.7660916)(14.264594,1.2660917)(13.584594,0.68609166)(12.904594,0.106091686)(11.884594,1.7660917)(11.444594,2.4660916)(11.004594,3.1660917)(8.444594,7.0460916)(8.304594,7.306092)(8.164594,7.5660915)(7.0045943,9.186091)(7.904594,9.626092)
\psdiamond[linewidth=0.02,dimen=outer](9.424594,1.1660917)(3.9,6.0)
\psline[linewidth=0.02cm](9.424594,7.1460915)(8.044594,9.166092)
\psline[linewidth=0.02cm](9.424594,-4.793908)(8.144594,-6.793908)
\psdots[dotsize=0.2](5.5245943,1.1460917)
\psdots[dotsize=0.2](9.264594,1.1460917)
\psdots[dotsize=0.2](13.284595,1.1460917)
\psdots[dotsize=0.2](6.864594,3.1660917)
\psdots[dotsize=0.2](8.144594,5.1460915)
\psdots[dotsize=0.2](10.684594,5.166092)
\psdots[dotsize=0.2](9.404594,7.1460915)
\psdots[dotsize=0.2](12.004594,3.1460917)
\psdots[dotsize=0.2](9.404594,-4.833908)
\psdots[dotsize=0.2](6.824594,-0.8539083)
\psdots[dotsize=0.2](8.124594,-2.8539083)
\psdots[dotsize=0.2](10.724594,-2.8539083)
\psdots[dotsize=0.2](12.004594,-0.8339083)
\psdots[dotsize=0.2](8.164594,-6.813908)
\psdots[dotsize=0.2](8.044594,9.146091)
\usefont{T1}{ppl}{m}{n}
\rput(10.029125,7.1560917){$\alpha_{4}$}
\usefont{T1}{ppl}{m}{n}
\rput(7.4191256,5.1560917){$2\alpha_{4}$}
\usefont{T1}{ppl}{m}{n}
\rput(11.989125,5.1760917){$\alpha_{3}+\alpha_{4}$}
\usefont{T1}{ppl}{m}{n}
\rput(13.269125,3.1560917){$\alpha_{2}+\alpha_{3}+\alpha_{4}$}
\usefont{T1}{ppl}{m}{n}
\rput(5.9591255,3.1760917){$\alpha_{3}+2\alpha_{4}$}
\usefont{T1}{ppl}{m}{n}
\rput(4.1091253,1.1560917){$2\alpha_{3}+2\alpha_{4}$}
\usefont{T1}{ppl}{m}{n}
\rput(9.579125,1.4160917){$\alpha_{2}+\alpha_{3}+2\alpha_{4}$}
\usefont{T1}{ppl}{m}{n}
\rput(15.209125,1.1560917){$\alpha_{1}+\alpha_{2}+\alpha_{3}+\alpha_{4}$}
\usefont{T1}{ppl}{m}{n}
\rput(5.1891255,-0.8239083){$\alpha_{2}+2\alpha_{3}+2\alpha_{4}$}
\usefont{T1}{ppl}{m}{n}
\rput(6.2791254,-2.8239083){$2\alpha_{2}+2\alpha_{3}+2\alpha_{4}$}
\usefont{T1}{ppl}{m}{n}
\rput(13.959126,-0.8239083){$\alpha_{1}+\alpha_{2}+\alpha_{3}+2\alpha_{4}$}
\usefont{T1}{ppl}{m}{n}
\rput(12.809125,-2.8839083){$\alpha_{1}+\alpha_{2}+2\alpha_{3}+2\alpha_{4}$}
\usefont{T1}{ppl}{m}{n}
\rput(11.619125,-4.8839083){$\alpha_{1}+2\alpha_{2}+2\alpha_{3}+2\alpha_{4}$}
\usefont{T1}{ppl}{m}{n}
\rput(10.649125,-6.7639084){$2\alpha_{1}+2\alpha_{2}+2\alpha_{3}+2\alpha_{4}$}
\usefont{T1}{ptm}{m}{n}
\rput(8.129437,-6.8039083){\color{red}\pscirclebox[linewidth=0.02,fillstyle=solid,fillcolor=red]{5}}
\usefont{T1}{ptm}{m}{n}
\rput(5.5110006,1.1160917){\color{red}\pscirclebox[linewidth=0.02,fillstyle=solid,fillcolor=red]{3}}
\usefont{T1}{ptm}{m}{n}
\rput(8.128032,5.1360917){\color{red}\pscirclebox[linewidth=0.02,fillstyle=solid,fillcolor=red]{2}}
\usefont{T1}{ptm}{m}{n}
\rput(8.022719,9.136091){\color{red}\pscirclebox[linewidth=0.02,fillstyle=solid,fillcolor=red]{1}}
\usefont{T1}{ptm}{m}{n}
\rput(8.120219,-2.8439083){\color{red}\pscirclebox[linewidth=0.02,fillstyle=solid,fillcolor=red]{4}}
\end{pspicture}
}

\caption{\label{fivecolourfigure}This figure describes the sets of weights of the ${\bf \overline{15}}$  that can be reached by an infinitesimal $\text{SL}(5,\mathbb{R})$ transformation starting from any of the five longest weights. Sets of weights associated to different longest weights correspond to different colours. As it can be seen from the figure, any short weight belongs to two sets, which means that one associates to it a 2-conjunction stabiliser. 
For each weight we write  its distance  from the highest weight in terms of simple roots. 
}
\end{figure}

We now consider the multiple-charge orbits of the ${\bf \overline{15}}$. In 
Fig.~\ref{fivecolourfigure} we show all the weights that can be reached by an infinitesimal $\text{SL}(5,\mathbb{R})$ transformation starting from any of the five long weights. We note that one can never reach a long weight starting from another one\,\footnote{Again, this is easy to understand in components because one cannot rotate for instance $T^{11}$ to any $T^{MM}$ with $M \neq 1$ with infinitesimal transformations.}. 
Each short weight $T^{MN}$, $M\neq N$, is connected infinitesimally to two long weights, $T^{MM}$ and $T^{NN}$. This implies that there is a conjunction stabiliser in the orbit of the bound state containing the branes corresponding to the charges $T^{MM}$ and $T^{NN}$. 
There are no $n$-conjunction stabilisers with $n>2$ because there are no weights that are connected to $n$ long weights for $n>2$.  The 2-charge orbit is
\begin{equation}
 \dfrac{\text{SL}(5,\mathbb{R})}{(\text{SL}(3,\mathbb{R})\times \text{SO}(1))\ltimes( T^{\bf 3}\times T^{\bf 3})} \ ,
\end{equation}
the three-charge orbit is
\begin{equation}
 \dfrac{\text{SL}(5,\mathbb{R})}{(\text{SL}(2,\mathbb{R})\times \text{SO}(3))\ltimes T^{\bf(3,2)}} \ ,
\end{equation}
while the four and five-charge orbits are respectively
\begin{equation}
 \dfrac{\text{SL}(5,\mathbb{R})}{\text{SO}(3)\ltimes T^{\bf 4}}
\end{equation}
and
\begin{equation}
 \dfrac{\text{SL}(5,\mathbb{R})}{\text{SO}(5)}.
\end{equation}
In particular, one can see that the stabilisers of the 5-charge orbit are the generators $E_\alpha - E_{-\alpha}$ for all the positive roots of $\text{SL}(5,\mathbb{R})$, and
the Cartan generators and the simple-root vectors of $\text{SO}(5)$ can be written as
\begin{eqnarray}
&&H_{\beta_{1}}=iE_{\alpha_{2}+\alpha_{3}}-iE_{-\alpha_{2}-\alpha_{3}}-iE_{\alpha_{3}+\alpha_{4}}
+iE_{-\alpha_{3}-\alpha_{4}}\,,\nonumber \\[.2truecm]
&&H_{\beta_{2}}=2iE_{\alpha_{3}+\alpha_{4}}-2iE_{-\alpha_{3}-\alpha_{4}}\,,\nonumber \\[.2truecm]
& &E_{\beta_{1}}=\frac{1}{4}\bigl(E_{\alpha_{2}}-E_{-\alpha_{2}}-iE_{\alpha_{2}+\alpha_{3}+\alpha_{4}}+
iE_{-\alpha_{2}-\alpha_{3}-\alpha_{4}}+E_{\alpha_{4}}-E_{-\alpha_{4}}-iE_{\alpha_{3}}+iE_{-\alpha_{3}}\bigr) \ ,\nonumber\\
&& E_{\beta_{2}}=\frac{\sqrt{2}}{4}\bigl(iE_{\alpha_{1}+\alpha_{2}}-iE_{-\alpha_{1}-\alpha_{2}}-E_{\alpha_{1}+\alpha_{2}
+\alpha_{3}+\alpha_{4}}+E_{-\alpha_{1}-\alpha_{2}-\alpha_{3}-\alpha_{4}}\bigr)\,,
\end{eqnarray}
where $\beta_1$ and $\beta_2$ are the simple roots of $\text{SO}(5)$.

We next analyse the invariants. The highest-weight (single-charge) orbit is defined by the constraint
  \begin{equation}
   T^{M_1 N_1} T^{M_2 N_2} \epsilon_{N_1 N_2 ...N_5}=0 \quad .
\end{equation}
 When this quantity is instead non-vanishing, but
 \begin{equation}
   T^{M_1 N_1} T^{M_2 N_2} T^{M_3 N_3} \epsilon_{N_1 N_2 ...N_5}=0 \quad ,\label{2chargeorbit15}
\end{equation}
one obtains the 2-charge orbit. Proceeding this way, one arrives at a five-charge orbit, for which the quantity
\begin{equation}
   T^{M_1 N_1} T^{M_2 N_2} T^{M_3 N_3} T^{M_4 N_4} T^{M_5 N_5} \epsilon_{N_1 N_2 ...N_5}
\end{equation}
is non-vanishing.

We finally consider the supersymmetry. The projection of the brane charge $T^{MN}$ on the singlet central charge $Q$ is (see section 3)
  \begin{equation}
T^{MN} \rightarrow \delta^{MN} Q \ ,
\end{equation}
which means that all the different constraints on the charges that define the five different orbits are all projected on the same $\text{SO}(5)$  epsilon symbol. This means that all these brane configurations preserve the same amount of supersymmetry.

The 7D example discussed above  can be generalised to other representations and other dimensions. In general we expect that if different brane orbits correspond to invariants that  lead to the same central charge constraints when projected on the R-symmetry, these brane configurations all preserve the same amount of supersymmetry. From Table  \ref{nonstandardcentralcharge} one can determine all these configurations in general. We hope to report on this in more detail in the near future.

\section{Conclusions}

In this work we studied several properties of branes in string theory with 32 supercharges from a purely group-theoretical point of view. We contrasted the
branes with three or more transverse directions, which we called ``standard'' branes, with the branes which have
two or less transverse directions, which we denominated ``non-standard'' branes. More specifically, we
called them ``defect'' branes (two transverse directions), domain walls (one transverse direction) and space-filling branes (no transverse direction).

We focussed on three distinct brane properties. First, we showed that the half-super\-sym\-me\-tric branes, both standard and non-standard ones, always correspond to the longest weights of the U-duality representation these branes belong to. It turns out that the
standard branes always occur in U-duality representations where all weights are longest weights. This explains why for standard branes the dimension of the U-duality representation equals the number of half-supersymmetric branes. In contrast,
the non-standard branes always occur in U-duality representations with different lengths of weights. This is why the number of half-supersymmetric non-standard branes is  always less than the dimension of the U-duality representation to which they belong. Using this simple group-theoretical characterization we calculated the number of half-supersymmetric non-standard branes, reproducing the results of
\cite{Bergshoeff:2011qk,Bergshoeff:2012ex,Kleinschmidt:2011vu}.
For defect branes the number is given by ${\rm dim }\,G - {\rm rank}\,
 G$ where $G$ is the U-duality group.
The answer for the  domain walls and space-filling branes can be found in Table \ref{dominantweightsofnonstandardbranes}.

We next studied the BPS properties of the standard and non-standard branes. Using a decomposition
of the U-duality representation of the brane charges into representations of the R-symmetry of the central charges we found a second
crucial difference between standard and non-standard branes. Whereas for standard branes for each BPS condition there is a
unique brane, we find that different non-standard branes may satisfy the same BPS condition. We calculated the degeneracy of these BPS conditions for all non-standard branes in different dimensions. The result can be found in Table
\ref{nonstandardcentralcharge}.

We finally discussed the standard and non-standard brane orbits. Our results on the multi-charge non-standard-brane orbits are new. We discussed the invariants that characterize these orbits and found that for non-standard branes different invariants
may project onto the same central charge showing that different brane configurations may preserve the same supersymmetry.

In our discussion the length of the weights of the  representations of the U-duality group $G$ played an important role. In particular, the longest weights were associated to the half-supersymmetric branes.
In \cite{Kleinschmidt:2011vu} the same counting of the half-supersymmetric branes was obtained using a different method, based upon the counting of the real roots of the very extended Kac-Moody algebra $E_{11}$. Considering the longest weights of the U-duality representations
 can indeed be translated to taking the real roots within the very extended Kac-Moody algebra $E_{11}$. A relation between the squared length of the $E_{11}$ roots and the squared length of the weights of the U-duality representations was given in the Appendix of \cite{Riccioni:2008jz}, based on the analysis of \cite{West:2004kb}. The relation consists in  writing down the expression of $\alpha^2$ for an $E_{11}$ root and  decomposing it in terms of the weights of the subalgebra $\text{SL}(D,\mathbb{R}) \times E_{11-D}$, where $E_{11-D}$ is the U-duality group $G$. One restricts the attention to the form fields, i.e.~to the completely antisymmetric representations of $\text{SL}(D,\mathbb{R})$. All these representations have only one dominant weight, which means that all the components give the same contribution to $\alpha^2$.
On the other hand, the representations of $E_{11-D}$ are decomposed in longest weights, next-to-longest weights, etc.   The difference between the squared length of the longest weights and the next-to-longest weights is equal to 2, which is the squared length of the roots, as noticed in section 2\,\footnote{Here we have normalised the squared length of the simple roots to 2 for simplicity.}.  On the other hand, the roots of $E_{11}$ have squared length $\alpha^2 =2,0,-2,-4,...$.
 This implies that the relation between the lengths of the weights of the U-duality representation and the lengths of the roots of the very extended Kac-Moody algebra $E_{11-D}$ is as follows:
\begin{equation}\label{relation}
\begin{array}{ccc}
\text{weights of}\ G && \text{roots of}\ E_{11-D}\\[.3truecm]
\text{longest}&& \alpha^2=2\\
\text{next-to-longest}&&\alpha^2=0\\
\text{next-to-next-to-longest}&&\alpha^2=\text{-}2\\
\vdots&& \vdots\end{array}
\end{equation}
This relation holds for the highest-dimensional representation. For a given form, smaller representations, whose highest weight coincides with one of the dominant weights of the highest-dimensional representation, also occur. For these fields the value of $\alpha^2$ is given by eq. \eqref{relation} where one has to pick the dominant weight of the highest-dimensional representation that has the same length as the highest weight of the lower-dimensional representation. These fields therefore have $\alpha^2<2$ and are not associated to branes.

Knowing that the longest weights of the U-duality representation correspond to the half-supersymmetric branes, it is natural to consider also the interpretation of the shorter weights. The first ones
 to consider are the next-to-longest weights,  corresponding to the $\alpha^2=0$ roots of $E_{11}$.
In the case of the 7-branes of IIB, the short weight is the Cartan of $\text{SL}(2,\mathbb{R})$, and a charge in the Cartan corresponds to a bound state of the D7-brane and its S-dual. This can be easily understood in terms of invariants. Given the charge $T^{\alpha\beta}$ in the ${\bf 3}$, the orbits are defined by the value of the invariant
\begin{equation}
T^{\alpha \beta} T^{\gamma \delta} \epsilon_{\alpha \gamma} \epsilon_{\beta \delta} \quad .
\end{equation} 
This quantity is vanishing for a highest-weight orbit, i.e. a single-charge orbit, corresponding to the charge $T^{11}$ or $T^{22}$, while it is non-vanishing for the charge $T^{11} + T^{22}$, corresponding to a bound state, as well as for  the charge $T^{12}$, which thus belongs to the same orbit as the bound state. A similar conclusion can be reached for the case of the ${\bf \overline{15}}$ of $\text{SL}(5,\mathbb{R})$ analysed in Section 4.2. Any charge $T^{MN}$, with $M \neq N$, satisfies the constraint \eqref{2chargeorbit15}, and thus corresponds to a 2-charge state. One could reach similar conclusions in all the other cases. It would be interesting to compare such an  analysis with  the work of \cite{West:2004st}, \cite{Englert:2004it}   or with the more recent work of \cite{Cook:2011ir}.

One of the results of our investigations is that lower-dimensional string theory contains many non-standard branes, much
more than the standard ones. It is natural to ask whether there are any applications of these branes. For a
recent application in the context of black holes, see \cite{deBoer:2012ma}.  As shown in \cite{Bergshoeff:2011mh,Bergshoeff:2011ee,Bergshoeff:2012ex}, the branes of the ten-dimensional theory satisfy generalised wrapping rules when compactified on tori. In the case of the fundamental string, these wrapping rules are a manifestation of the stringy doubled geometry discussed in
\cite{Hull:2004in}. It would be interesting to see whether a similar geometric interpretation can be given for the wrapping rules of the other branes, as well as for the branes, among those listed in this paper, that do not follow from any wrapping rule from the branes of the ten-dimensional theory.

It is natural to extend our work to the the branes of half-maximal supergravities or the supergravity theories with even less supersymmetry. The branes of the half-maximal supergravities have been obtained in \cite{Bergshoeff:2012jb} using the so-called `light-cone rule' derived in \cite{Bergshoeff:2011zk}. We expect that this rule can be translated  to general group-theoretic properties that can also be determined for the
 more complicated U-duality symmetries that occur in even less supersymmetric theories, exactly as we did for the case of maximally non-compact groups in this paper.
  We hope to report on progress in this direction  in the nearby future.

\vskip 1.5cm

\section*{Acknowledgements}

E.A.B. wishes to thank the University of Rome ``La Sapienza'' and INFN Sezione di Roma where part of this work was done for its hospitality. F.R. would like to thank A. Marrani for discussions.


\vskip 1.5cm



\begin{thebibliography}{99}

\bibitem{Bergshoeff:1987cm}
  E.~Bergshoeff, E.~Sezgin and P.~K.~Townsend,
  ``Supermembranes and Eleven-Dimensional Supergravity,''  Phys.\ Lett.\ B {\bf 189} (1987) 75.  


\bibitem{Polchinski:1995mt}
  J.~Polchinski,
  ``Dirichlet Branes and Ramond-Ramond charges,''  Phys.\ Rev.\ Lett.\  {\bf 75} (1995) 4724  [hep-th/9510017].  


\bibitem{Strominger:1996sh}
  A.~Strominger and C.~Vafa,
  ``Microscopic origin of the Bekenstein-Hawking entropy,''  Phys.\ Lett.\ B {\bf 379} (1996) 99  [hep-th/9601029].  

\bibitem{Maldacena:1997re}
  J.~M.~Maldacena,
  ``The Large N limit of superconformal field theories and supergravity,''  Adv.\ Theor.\ Math.\ Phys.\  {\bf 2} (1998) 231  [hep-th/9711200].  


\bibitem{Randall:1999ee}
  L.~Randall and R.~Sundrum,
  ``A Large mass hierarchy from a small extra dimension,''  Phys.\ Rev.\ Lett.\  {\bf 83} (1999) 3370  [hep-ph/9905221];
  {\sl ibid.}\~
  ``An Alternative to compactification,''  Phys.\ Rev.\ Lett.\  {\bf 83} (1999) 4690  [hep-th/9906064].  

\bibitem{de Azcarraga:1989gm}
  J.~A.~de Azcarraga, J.~P.~Gauntlett, J.~M.~Izquierdo and P.~K.~Townsend,
  ``Topological Extensions of the Supersymmetry Algebra for Extended Objects,''  Phys.\ Rev.\ Lett.\  {\bf 63} (1989) 2443.  

\bibitem{Greene:1989ya}
  B.~R.~Greene, A.~D.~Shapere, C.~Vafa and S.~-T.~Yau,
  ``Stringy Cosmic Strings and Noncompact Calabi-Yau Manifolds,''  Nucl.\ Phys.\ B {\bf 337} (1990) 1.  

\bibitem{Gibbons:1995vg}
  G.~W.~Gibbons, M.~B.~Green and M.~J.~Perry,
  ``Instantons and seven-branes in type IIB superstring theory,''  Phys.\ Lett.\ B {\bf 370} (1996) 37  [hep-th/9511080].  


\bibitem{Vafa:1996xn}
  C.~Vafa,
  ``Evidence for F theory,''  Nucl.\ Phys.\ B {\bf 469} (1996) 403  [hep-th/9602022].  


\bibitem{Bergshoeff:2011se}
  E.~A.~Bergshoeff, T.~Ort\'\i n and F.~Riccioni,
  ``Defect Branes,''  Nucl.\ Phys.\ B {\bf 856} (2012) 210  [arXiv:1109.4484 [hep-th]].  


\bibitem{Bergshoeff:2010xc}
  E.~A.~Bergshoeff and F.~Riccioni,
  ``D-Brane Wess-Zumino Terms and U-Duality,''  JHEP {\bf 1011} (2010) 139  [arXiv:1009.4657 [hep-th]].  

\bibitem{Bergshoeff:2005ac}
  E.~A.~Bergshoeff, M.~de Roo, S.~F.~Kerstan and F.~Riccioni,
  ``IIB supergravity revisited,''  JHEP {\bf 0508} (2005) 098  [hep-th/0506013];  
  E.~A.~Bergshoeff, M.~de Roo, S.~F.~Kerstan, T.~Ort\'\i n and F.~Riccioni,
  ``IIA ten-forms and the gauge algebras of maximal supergravity theories,''  JHEP {\bf 0607} (2006) 018  [hep-th/0602280];  
  E.~A.~Bergshoeff, J.~Hartong, P.~S.~Howe, T.~Ort\'\i n and F.~Riccioni,
  ``IIA/IIB Supergravity and Ten-forms,''  JHEP {\bf 1005} (2010) 061  [arXiv:1004.1348 [hep-th]].  


\bibitem{Riccioni:2007au}
  F.~Riccioni and P.~C.~West,
  ``The E(11) origin of all maximal supergravities,''  JHEP {\bf 0707} (2007) 063  [arXiv:0705.0752 [hep-th]].  

\bibitem{Bergshoeff:2007qi}
  E.~A.~Bergshoeff, I.~De Baetselier and T.~A.~Nutma,
  ``E(11) and the embedding tensor,''  JHEP {\bf 0709} (2007) 047  [arXiv:0705.1304 [hep-th]].  




\bibitem{West:2001as}
  P.~C.~West,
  ``E(11) and M theory,''  Class.\ Quant.\ Grav.\  {\bf 18} (2001) 4443  [hep-th/0104081].  

\bibitem{deWit:2008ta}
  B.~de Wit, H.~Nicolai and H.~Samtleben,
  ``Gauged Supergravities, Tensor Hierarchies, and M-Theory,''  JHEP {\bf 0802} (2008) 044  [arXiv:0801.1294 [hep-th]].  

\bibitem{Bergshoeff:2011qk}
  E.~A.~Bergshoeff and F.~Riccioni,
  ``The D-brane U-scan,''  arXiv:1109.1725 [hep-th].  

\bibitem{Bergshoeff:2012ex}
  E.~A.~Bergshoeff, A.~Marrani and F.~Riccioni,
  ``Brane orbits,''  Nucl.\ Phys.\ B {\bf 861} (2012) 104  [arXiv:1201.5819 [hep-th]].  



\bibitem{Kleinschmidt:2011vu}
  A.~Kleinschmidt,
  ``Counting supersymmetric branes,''  JHEP {\bf 1110} (2011) 144  [arXiv:1109.2025 [hep-th]].  




\bibitem{Bergshoeff:2012pm}
  E.~A.~Bergshoeff, A.~Kleinschmidt and F.~Riccioni,
  ``Supersymmetric Domain Walls,''  Phys.\ Rev.\ D {\bf 86} (2012) 085043  [arXiv:1206.5697 [hep-th]].  

\bibitem{Bergshoeff:2012jb}
  E.~A.~Bergshoeff and F.~Riccioni,
  ``Heterotic wrapping rules,''  JHEP {\bf 1301} (2013) 005  [arXiv:1210.1422 [hep-th]].  



\bibitem{Lu:1997bg}
  H.~Lu, C.~N.~Pope and K.~S.~Stelle,
  ``Multiplet structures of BPS solitons,''  Class.\ Quant.\ Grav.\  {\bf 15} (1998) 537  [hep-th/9708109].  


\bibitem{Ferrara:1997ci}
  S.~Ferrara and J.~M.~Maldacena,
  ``Branes, central charges and U duality invariant BPS conditions,''  Class.\ Quant.\ Grav.\  {\bf 15} (1998) 749  [hep-th/9706097].  


\bibitem{Cahn:1985wk}
  R.~N.~Cahn,
  ``Semisimple Lie Algebras And Their Representations,''  Menlo Park, Usa: Benjamin/cummings ( 1984) 158 P. ( Frontiers In Physics, 59)


\bibitem{Bergshoeff:2006gs}
  E.~A.~Bergshoeff, M.~de Roo, S.~F.~Kerstan, T.~Ort\'\i n and F.~Riccioni,
  ``SL(2,R)-invariant IIB Brane Actions,''  JHEP {\bf 0702} (2007) 007  [hep-th/0611036].  

\bibitem{Ferrara:1997uz}
  S.~Ferrara and M.~Gunaydin,
  ``Orbits of exceptional groups, duality and BPS states in string theory,''  Int.\ J.\ Mod.\ Phys.\ A {\bf 13} (1998) 2075  [hep-th/9708025].  


\bibitem{Bossard:2009my}
  G.~Bossard and H.~Nicolai,
  ``Multi-black holes from nilpotent Lie algebra orbits,''
  Gen.\ Rel.\ Grav.\  {\bf 42} (2010) 509
  [arXiv:0906.1987 [hep-th]].

\bibitem{Borsten:2010aa}
  L.~Borsten, D.~Dahanayake, M.~J.~Duff, S.~Ferrara, A.~Marrani and W.~Rubens,
  ``Observations on Integral and Continuous U-duality Orbits in N=8 Supergravity,''  Class.\ Quant.\ Grav.\  {\bf 27} (2010) 185003  [arXiv:1002.4223 [hep-th]].  

\bibitem{Riccioni:2008jz}
  F.~Riccioni, A.~Van Proeyen and P.~C.~West,
  ``Real forms of very extended Kac-Moody algebras and theories with eight supersymmetries,''  JHEP {\bf 0805} (2008) 079  [arXiv:0801.2763 [hep-th]].  



\bibitem{West:2004kb}
  P.~C.~West,
  ``E(11) origin of brane charges and U-duality multiplets,''  JHEP {\bf 0408} (2004) 052  [hep-th/0406150].  

\bibitem{West:2004st}
  P.~C.~West,
  ``The IIA, IIB and eleven-dimensional theories and their common E(11) origin,''  Nucl.\ Phys.\ B {\bf 693} (2004) 76  [hep-th/0402140].  

\bibitem{Englert:2004it}
  F.~Englert and L.~Houart,
  ``G+++ invariant formulation of gravity and M-theories: Exact intersecting brane solutions,''  JHEP {\bf 0405} (2004) 059  [hep-th/0405082].  




\bibitem{Cook:2011ir}
  P.~P.~Cook,
  ``Bound States of String Theory and Beyond,''  JHEP {\bf 1203} (2012) 028  [arXiv:1109.6595 [hep-th]].  

\bibitem{deBoer:2012ma}
  J.~de Boer and M.~Shigemori,
  ``Exotic Branes in String Theory,''  arXiv:1209.6056 [hep-th].  





\bibitem{Bergshoeff:2011mh}
  E.~A.~Bergshoeff and F.~Riccioni,
  ``Dual doubled geometry,''  Phys.\ Lett.\ B {\bf 702} (2011) 281  [arXiv:1106.0212 [hep-th]].  


\bibitem{Bergshoeff:2011ee}
  E.~A.~Bergshoeff and F.~Riccioni,
  ``Branes and wrapping rules,''  Phys.\ Lett.\ B {\bf 704} (2011) 367  [arXiv:1108.5067 [hep-th]].  

\bibitem{Hull:2004in}
  C.~M.~Hull,
  ``A geometry for non-geometric string backgrounds,''
  JHEP {\bf 0510} (2005) 065
  [arXiv:hep-th/0406102];
  {\sl ibidem},
  ``Doubled geometry and T-folds,''
  JHEP {\bf 0707} (2007) 080
  [arXiv:hep-th/0605149];
  C.~M.~Hull and R.~A.~Reid-Edwards,
  ``Gauge Symmetry, T-Duality and Doubled Geometry,''
  JHEP {\bf 0808} (2008) 043
  [arXiv:0711.4818 [hep-th]].


\bibitem{Bergshoeff:2011zk}
  E.~A.~Bergshoeff and F.~Riccioni,
  ``String Solitons and T-duality,''  JHEP {\bf 1105} (2011) 131  [arXiv:1102.0934 [hep-th]].  






\end{thebibliography}
\end{document}